\documentclass[12pt,preprint]{aastex}
\usepackage{amsmath}
\usepackage{mathtools}
\newcommand{\etal}{et~al.\ }
\newcommand{\HI}{\hbox{{\rm H}\kern 0.1em{\sc i}}}
\newcommand{\HII}{\hbox{{\rm H}\kern 0.1em{\sc ii}}}
\newcommand{\CII}{\hbox{\rm[{C}\kern 0.1em{\sc ii}]}}
\newcommand{\Ha}{\hbox{{\rm H}\kern 0.1em$\alpha$}}
\newcommand{\msun}{${\rm M}_{\odot}$}

\newcommand{\persec}{s$^{-1}$}

\begin{document}
%\slugcomment{The Astrophysical Journal, accepted July 2013}
 
\shortauthors{KNIERMAN ET~AL.}
\shorttitle{Star Formation in Tidal Tails of NGC 2782}

\title{Tidal Tails of Minor Mergers II: Comparing Star Formation in the Tidal Tails of NGC 2782\footnotemark[1]}

\footnotetext[1]{Herschel is an ESA space observatory with science instruments provided by European-led Principal Investigator consortia and with important participation from NASA.}

\author{Karen A. Knierman}
\affil{School of Earth \& Space Exploration, Arizona State University,
550 E. Tyler Mall, Room PSF-686 (P.O. Box 871404),
TEMPE, AZ 85287-1404}
\email{karen.knierman@asu.edu}

\author{Paul Scowen}
\affil{School of Earth \& Space Exploration, Arizona State University,
550 E. Tyler Mall, Room PSF-686 (P.O. Box 871404),
TEMPE, AZ 85287-1404}
\email{paul.scowen@asu.edu}

\author{Todd Veach}
\affil{School of Earth \& Space Exploration, Arizona State University,
550 E. Tyler Mall, Room PSF-686 (P.O. Box 871404),
TEMPE, AZ 85287-1404}
\email{tveach@asu.edu}

\author{Christopher Groppi}
\affil{School of Earth \& Space Exploration, Arizona State University,
550 E. Tyler Mall, Room PSF-686 (P.O. Box 871404),
TEMPE, AZ 85287-1404}
\email{cgroppi@asu.edu}

\author{Brendan Mullan}
\affil{Department of Astronomy \& Astrophysics, Penn State University, 
525 Davey Lab, University Park, PA }
\email{mullan@astro.psu.edu}

\author{Iraklis Konstantopoulos}
\affil{Australian Astronomical Observatory, P.O. Box 915, North Ryde NSW 1670}
\email{ iraklis@aao.gov.au}

\author{Patricia M. Knezek}
\affil{WIYN Consortium, Inc., 
950 N. Cherry Avenue,
Tucson, AZ  85719}
\email{pknezek@noao.edu}

\author{Jane Charlton}
\affil{Department of Astronomy \& Astrophysics, Penn State University, 
525 Davey Lab, University Park, PA }

\begin{abstract}
The peculiar spiral NGC 2782 is the
result of a minor merger with a mass ratio $\sim4:1$
occurring $\sim200$ Myr ago. This merger produced a molecular and {\HI} rich,
optically bright Eastern tail and an {\HI}-rich, optically faint
Western tail.   Non-detection of CO in the Western Tail by \citet{braine} suggested that
star formation had not yet begun.  However, deep $UBVR$ and {\Ha} narrowband images 
show evidence of recent star formation in the Western tail, though it lacks massive star clusters and cluster complexes.  Using Herschel PACS spectroscopy, we discover 158$\mu$m {\CII} emission at the location of the three most luminous {\Ha} sources in the Eastern tail, but not at the location of the even brighter {\Ha} source in the Western tail.  
The Western tail is found to have a normal star formation efficiency (SFE), but the Eastern tail has a low SFE.  
The lack of  CO and {\CII} emission suggests the Western tail {\HII} region may have a low carbon abundance and be undergoing its first star formation.   
The Western tail is more efficient at forming stars, but lacks massive clusters.  We propose that the low SFE in the Eastern tail may be due to its formation as a splash region where gas heating is important even though it has sufficient molecular and neutral gas to make massive star clusters.  The Western tail, which has lower gas surface density and does not form high mass star clusters, is a tidally formed region where gravitational compression likely enhances star formation.  
%ADD SENTENCE ABOUT CII in E tail

\end{abstract}


\section{Introduction}
Major mergers (e.g., two equal mass spiral galaxies) can form in their debris ranging from large 
 tidal dwarf galaxies (TDGs) down to star clusters
 \citep[e.g.,][]{duc,wfdf02,knierman,mullan}.  However, the
spectacular displays of major mergers  \citep[e.g., NGC
4038/9 ``The Antennae'';][]{whitmore99} are relatively rare compared to minor mergers
(in this work, defined to be between a dwarf galaxy and spiral galaxy with a mass ratio of $<
0.3$).  This work aims to
understand how minor mergers shape galactic structure and also to 
examine star formation in gas that may be marginally stable. 

Observations of young star clusters in tidal debris show varied results, with some tails forming many star clusters, while 
others only have a few star clusters scattered along the tail or only hosted in the tidal dwarf galaxy.  Studies have shown a large amount of neutral hydrogen in tidal debris \citep{smith94, hibbard, aparna}, but only certain regions with higher gas densities host molecular gas \citep{smith99, braine}.  Previous results \citep{boquien, knierman12} have also shown that star formation in tidal debris does not follow the Kennicutt-Schmidt law.  
This work aims to compare star formation and gas properties on local and global scales using multiwavelength observations of two tidal tails with 
different properties in the same system.

The peculiar spiral, NGC 2782, is at a distance of $39.5$\footnote{From NED, corrected for Virgo, Great Attractor, and Shapley, which we will use for the duration of this paper.}
 Mpc \citep{mould}.  in the RC3, it is classified as SABa(rs) pec 
since it has a disturbed center with bright arcs.  A starburst is occurring in the central regions \citep{devereux}.  NGC 2782 has two tidal tails: an {\HI}-rich, optically faint
Western tail extending 5\arcmin \ to the northwest and an Eastern tail which
has a concentration of {\HI} and CO at its base, but a gas-poor
optically bright knot 2.7\arcmin \ from the center \citep[see][]{smith94}.  
A tidal dwarf galaxy candidate (TDGC) was discovered by \citet{yoshida}
in the Eastern tail near the main body of the galaxy. Their optical spectrum confirms its association with NGC 2782 and  indicates that it may be metal poor, similar to other TDGs \citep{duc}.     
CO is not detected at the 
location of massive {\HI} clumps in the Western tail which led \citet{braine} to
suggest that the {\HI} in the Western tail of NGC 2782 is not gravitationally bound and ``has presumably
not had time to condense into H$_2$ and for star formation to begin."

To determine the merger age and geometry,
\citet{smith94} constructed a restricted 3-body dynamical model of NGC 2782.   This model reproduces
the morphology and {\HI} velocities which indicates that NGC 2782 may be the result
of a merger between a large disk galaxy and a lower mass disk galaxy
with a mass ratio of $\sim0.25$ occurring $\sim200$ Myr ago.  However,
this model does not include gas dynamics or self-gravity of the
particles which could change the results.  Further simulations have
not been done to test this merger scenario.  Merger age can
also be inferred by using the maximum tail length (50 kpc) and the
disk rotation speed (150 km s$^{-1}$; \citet{smith94}) (see Section
3.1 of \citet{knierman}).  For NGC 2782, we infer a merger age of 50
kpc/150 km s$^{-1} = 300$ Myr which is close to the age from
\citet{smith94}.

\citet{wehnerthesis} also studied the tidal debris of NGC 2782 with deep,
wide-field imaging and found that the debris in the Eastern and
Western plumes has colors both consistent with each other and bluer
than the main disk of NGC 2782, suggesting that perhaps the two tails
formed from the same dwarf companion that passed through or along the disk of NGC
2782 and was destroyed in the process.  However, by examining
the $m_{\scriptsize\HI}/L_B$ ratios for the stellar and gaseous debris,
\citet{wehnerthesis} concludes that it is unlikely that all the
gaseous debris in the Western plume originated in the smaller
companion, as this would require the excessively large $m_{\HI}/L_B$ ratio
of 6.4.  More likely, a significant amount of the gaseous debris
originated in the gaseous disk of the main galaxy.

NGC 2782 also has a well-behaved exponential disk at intermediate
radii.  Smith et al. (1994) find an $R^{1/4}$ profile within the
innermost arcminute, and \citet{wehnerthesis} finds that the $R^{1/4}$
profile reemerges at higher radii, consistent with the idea that the
outer stellar plumes are debris resulting from a minor merger.
 \citet{mullan} in their $V$ and $I$ band \emph{Hubble Space Telescope}/WFPC2 survey of tidal tails find 87 star cluster candidates in the Eastern tail of NGC 2782 and 10 candidates in the Western tail. 

We obtained deep optical broadband and {\Ha} images combined with new {\CII}  and CO observations and published {\HI} and CO
observations to compare local and global star formation, as determined by several different tracers, between the tidal tails in NGC 2782.  Section 2 
contains the observations and calibrations.  Section 3 presents the results.  In Section 4, we discuss the possible reasons for the differences between the tidal tails.  

\section{Observations and Reductions}
We examine different tracers of star formation in the tidal tails of NGC 2782.  First, we examine both tails for evidence of young star clusters using ground-based $UBVR$ images and HST/WFPC2 images to identify isolated star clusters or star cluster complexes.  The HST/WFPC2 images are from the Cycle 16 program 11134 (P.I. K. Knierman) and published in \citet{mullan}.  Next, we identify young star forming regions using {\Ha} and {\CII}.
Then we examine the amount of gas available for star formation through new and published CO observations and previous {\HI} observations.  
 
\subsection{Optical images}
\begin{figure}
\includegraphics[width=\textwidth]{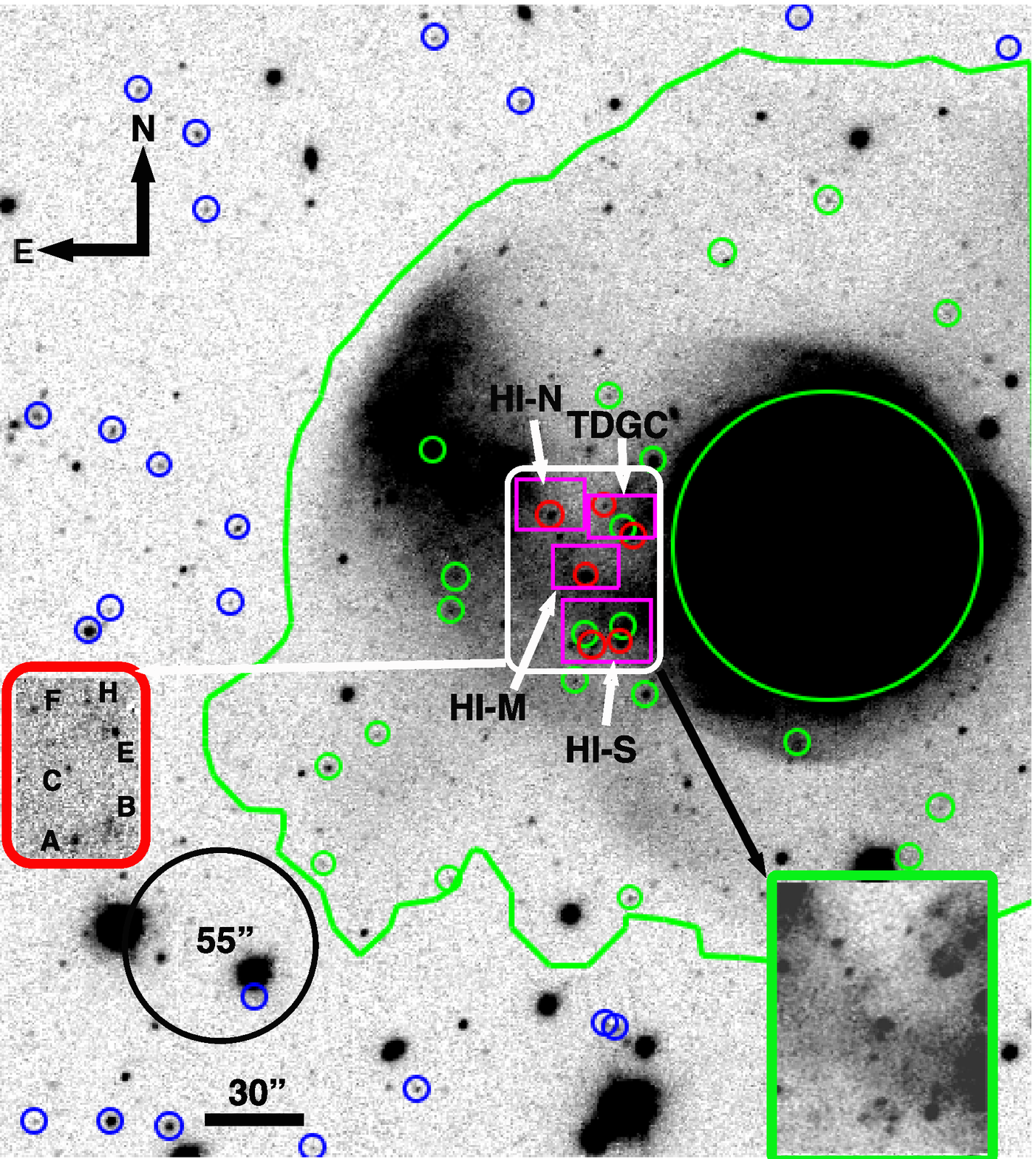}
\caption{$V$ image of Eastern tail taken at the \emph{VATT} 1.8m with final star cluster candidates marked by small circles.  
Green indicates ``in tail", while Blue indicates ``out of tail".  The large green
circle indicates the area defined as the central region of the
galaxy. Red circles mark the locations of {\Ha} sources.  Magenta boxes
mark the locations of massive {\HI} clouds \citep{smith94}. The inset outlined in red shows the {\Ha} emission from 
the area indicated by the white box. The inset image outlined in green shows an enlargement of the region of the $V$ image indicated by the white box.  The black circle indicates the beam size of the CO(1-0) observations with the ARO Kitt Peak 12 meter telescope.}
\label{fig:VimageE}
\end{figure}

\begin{figure}
\includegraphics[width=\textwidth]{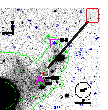}
\caption{$V$ image of Western tail taken at the \emph{VATT} 1.8m with final star cluster candidates marked by small circles.  
Green indicates ``in tail", while Blue indicates ``out of tail".  
The large green circle indicates the area defined as the central region of the galaxy.
Red circles mark the location of the {\Ha} source, W235 \citep{knierman12}.  Magenta boxes mark the
locations of massive {\HI} clouds \citep{smith94}.  The crosses denote the locations where \citet{smith99} looked for CO.  The blue cross was reobserved with more sensitive observations by \citet{braine}.  The inset shows the {\Ha} emission from the area indicated by the white box. The black circle indicates the beam size of the CO(1-0) observations with the ARO Kitt Peak 12 meter telescope.}
\label{fig:VimageW}
\end{figure}

Images in $UBVR$ and {\Ha} were taken with the Loral 2K CCD imager at the Lennon 1.8m \emph{Vatican
Advanced Technology Telescope} (\emph{VATT}) on Mount Graham, Arizona
(see Table 1 for a log of observations and Figures~\ref{fig:VimageE} and \ref{fig:VimageW}).  This imager has
a 6.4\arcmin \ field of view with 0.42{\arcsec} \ per pixel.  
Images were reduced using standard IRAF\footnote{IRAF is distributed by the National Optical Astronomy
Observatory, which is operated by the Association of Universities for Research in Astronomy, Inc., under cooperative 
agreement with the National Science Foundation.} tasks.  

%\clearpage
\begin{deluxetable}{lcccc}
\tablecaption{Optical Observations of NGC 2782}
\label{tab:n2782obs}
\tablewidth{6in}
\tablewidth{0pt}
\tablehead{\colhead{Tail} & \colhead{Date} & \colhead{Filter} & \colhead{Exp. Time} &\colhead{Photometric?} \\
\colhead{} & \colhead{} & \colhead{} & \colhead{sec} & \colhead{} }
\startdata
Eastern Tail & Feb. 15, 2004 & U & $12\times300$ & yes\\
 & Feb. 17, 2004 & U & $3\times900$ & no\\
%& Nov. 30, 2005 & U & ? & yes\\
\hline
& Feb. 15, 2004 & B & $9\times300$ & yes\\
\hline
& Feb. 17, 2004 & V & $4\times900$ & no\\
& May 12, 2005 & V & $2\times300$ & yes\\
\hline
& Feb. 17, 2004 & R & $3\times900$ & no\\
& May 12, 2005 & R & $2\times300$ & yes\\
\hline
& May 12, 2005 & H$\alpha$ & $3\times1200$ & yes\\
\hline
\hline
Western Tail& Feb. 18, 2004 & U & $8\times900$ & no\\
& May 22, 2004 & U & $2\times600$  & yes\\
\hline
& Feb. 16, 2004 & B & $4\times900$ & no\\
& May 22, 2004 & B & $2\times300$ & yes\\
\hline
& Feb. 17, 2004 & V & $4\times900$ & no\\
& May 22, 2004 & V & $2\times300$ & yes\\
\hline
& Feb. 17, 2004 & R & $3\times900$ & no\\
& May 22, 2004 & R & $2\times300$ & yes\\
& Oct. 3, 2004 & R & $3\times150$  & yes\\
& Oct. 4, 2004 & R & $2\times300$ & yes\\
\hline
& Oct. 3, 2004 & H$\alpha$ & $3\times900$ & yes\\
& Oct. 4, 2004 & H$\alpha$ & $2\times1200$ & yes\\
& Dec. 6, 2004 & H$\alpha$ & $3\times900$ & no\\

\hline
\hline

\enddata
\end{deluxetable}

\subsubsection{Source Detection and Photometry}\label{photometry}
To select sources, IRAF-DAOFIND was used with a threshold of $4\sigma$
where $\sigma$ is the standard deviation of the background.  
The
standard deviation of the background was found by averaging 
several regions around the image, away from the varying
background of the tail.  Sources with $S/N > 3.0$ were
retained for photometry due to the smoothness of the ground-based imaging.

Aperture photometry was performed on these sources using the
PHOT task in the IRAF-APPHOT package.  The radii of the object aperture,
the inner boundary of the background annulus, and the outer boundary
were 8, 13, and 18 pixels, respectively (3.4/5.5/7.6 arcsec).  We
retained only sources that were detected in all bands ($UBVR$) and with
magnitude errors $<0.2$.  Photometric zero-points were obtained
using Landolt standard stars taken on photometric nights.  The
foreground extinction from the Galaxy was corrected using the $A_B$
values from \citet{Schlegel98} and the reddening curve from
\citet{mathis90}.

\subsubsection{Selection of In Tail and Out of Tail Regions}
To obtain a population of background objects we identified ``in tail''
and ``out of tail'' regions.  
Previous studies have used various techniques to determine tail boundaries, including
location of the {\HI} contours \citep{knierman}, surface brightness
contours at a specific surface brightness ($\mu$) \citep{schombert}, and visual inspection
\citep{bastian}.  For this study, a combination of surface brightness
and contour plotting methods was used.  There are two areas to define
for tidal debris: the inner regions of the galaxy must be
excluded to prevent contamination from star clusters formed in the
dense inner starburst, and the outer extent of the debris region where
the debris fades into the general sky background.

To determine where the disk of the galaxy ends and the debris begins, we considered 
the use of the standard $D_{25}$ system or the isophotal diameter at the limiting surface 
brightness of 25 B mag arcsec$^{-2}$.  
The major and minor axes and the position angle of the $D_{25}$ parameters
were taken from LEDA\footnote{HyperLeda database (http://leda.univ-lyon1.fr) \cite{leda}}, and assuming an ellipse, 
plotted on the image.
%In general, the disk of the galaxy extended beyond this ellipse.
%Then, each $V$ image had a contour at $\mu_V = 25$ mag arcsec$^{-2}$
%plotted on it.  This contour was often at a larger extent than the $D_{25}$
%taken from LEDA.  Since the LEDA data include the RC3, $D_{25}$ in RC3 is
%often off from $\mu_V = 25$ by 0.4 mag (R. Jansen, private comm.).
%And often, the contour at $\mu_V = 25$ mag arcsec$^{-2}$ extended into
%the debris region.  
Next, logarithmically spaced surface brightness contours were
plotted on the image to determine where the regularity of the central
galaxy region departed into the asymmetry of the debris region.  The
boundary between the central region of the galaxy and the tidal debris 
was then defined to be a combination of the $D_{25}$ ellipse
and the last regularly shaped contour.  %See Figures~\ref{fig:VimageE}-\ref{fig:VimageW} for examples.  

To determine the divide between the debris region and the general
background, again, a combination of contour plots and surface
brightness limits was used.  We first plot the Holmberg radius \citep{holmberg}, the radius at
which the surface brightness in a blue filter reaches 26.5 mag
arcsec$^{-2}$.  This contour at $\mu_B = 26.5$ mag
arcsec$^{-2}$, followed not the central regions of this galaxy, but the outer boundary of the tidal tails and
debris regions.  Holmberg found that the radius at which $\mu_B =
26.5$ mag arcsec$^{-2}$ included the outermost {\HII} or star forming
regions in a galaxy, however, recent work shows this to not always be the case \citep{elmhunter}.  
It is, therefore, not surprising that the
Holmberg radius in this merger includes the tidal debris
with its star formation.  To ensure inclusion of all tidal debris, 
we plot logarithmically spaced surface brightness contours on the images. 
At the edges of the tidal debris, there was a rapid decline in surface
brightness denoted by the small spacing of several contours followed by 
2 surface brightness contours with larger spacing before the tidal debris faded into the background.
The outermost contour that enclosed the tail region was chosen to
represent the tidal tail.  Upon inspection, this contour followed
approximately the Holmberg contour, however, where they differed
greatly, the contour chosen by contour spacing was preferred. 
Figures~\ref{fig:VimageE} and \ref{fig:VimageW} indicate the tidal tail regions
selected.

\subsubsection{Completeness Tests}

\begin{figure}
\plottwo{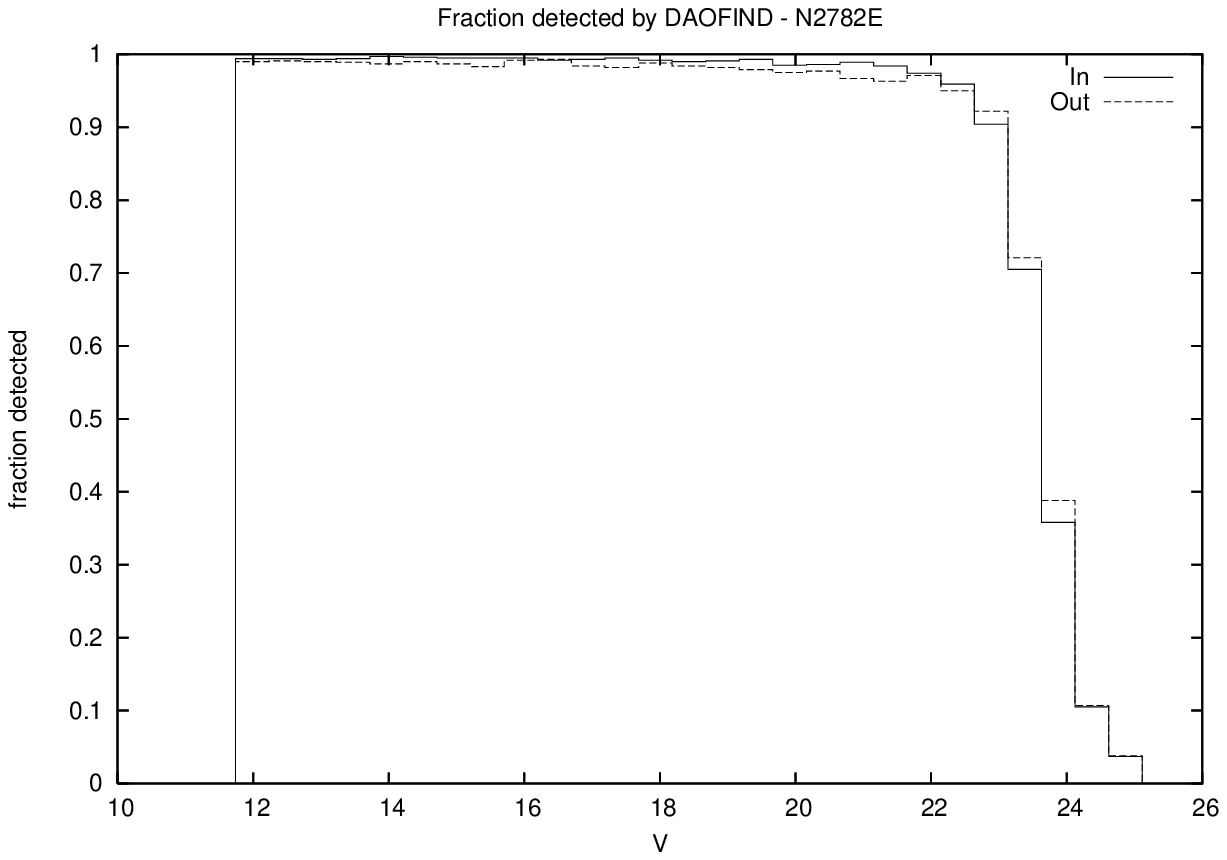}{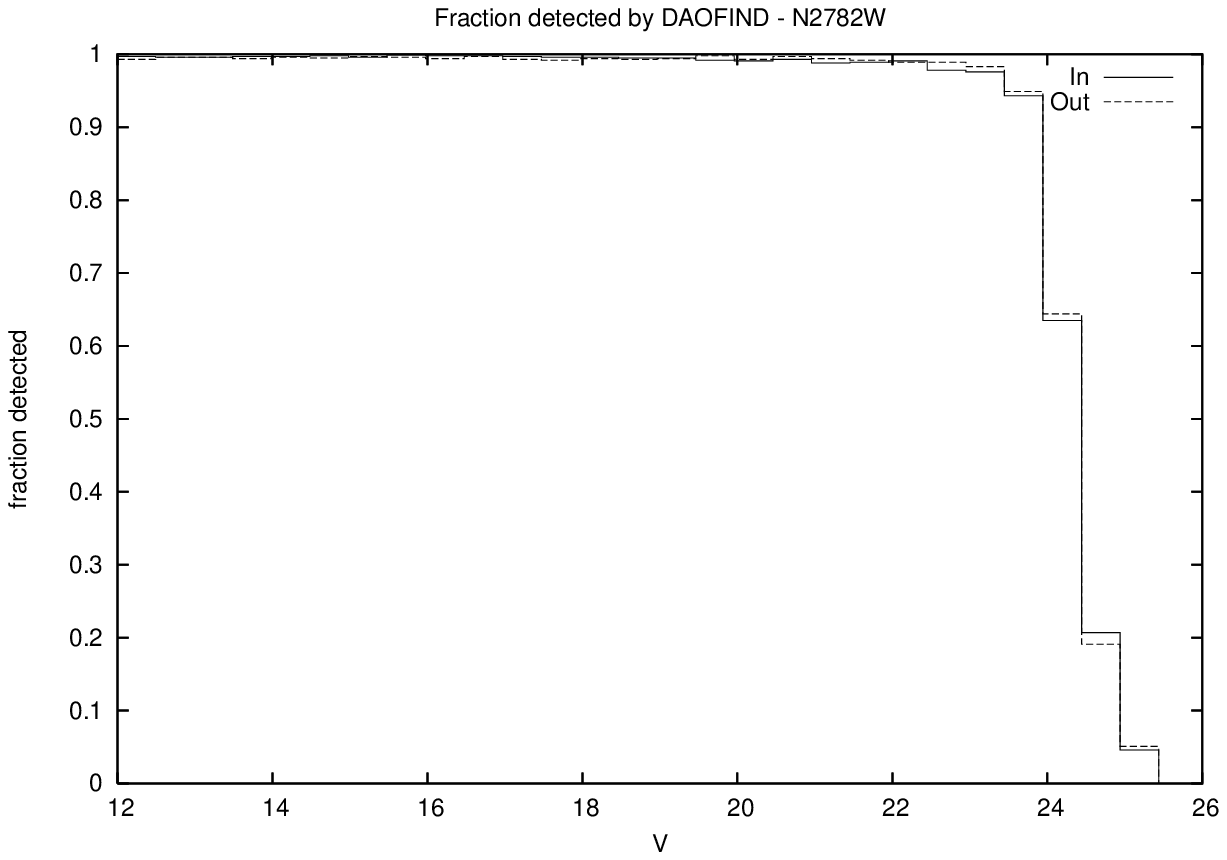}
\caption{For Eastern Tail (left) and Western Tail (right), the fraction of artificial stars recovered by DAOFIND in the $V$ band.}
\label{fig:complE}
\end{figure}

Completeness tests were conducted on the images in each band. First, a
PSF was constructed using 6-10 bright, isolated stars in each image
with the PSF task.  Then ADDSTAR was used to add 100 stars randomly
distributed across the image.  Stars were added in 0.5 mag bins
spanning the range from 13-26.5 instrumental magnitudes (e.g., an
image with 100 stars added with magnitudes between 13 and 13.5, a
second image with 100 stars added with magnitudes between 13.5 and
14).  Sources were detected and photometry performed as described in Section~\ref{photometry}.
This procedure was run
50 times per image to produce a total of 5000 stars added randomly to
each image.  
As seen in Figure~\ref{fig:complE}, the completeness limit for the Eastern Tail is
$V\sim22.9$ ($M_V\sim-10.1$) and for the Western Tail is $V\sim23.7$
($M_V\sim-9.3$).  The higher background of the Eastern tail may account for the completeness
limit at a brighter magnitude.

\subsubsection{Final Star Cluster Candidates}
To select the final sample of star cluster candidates, we used the
3DEF method \citep{bik}, a three dimensional maximum likelihood code
which fits the spectral energy distribution (SED) of a cluster to broad band
colors, to fit cluster ages, masses, and extinctions  (number of data points = 4, parameters in model = 3).  
For each
cluster, the 3DEF method uses a grid of simple stellar population
(SSP) models of ages between 1 Myr and 10 Gyr, extinctions ranging
from $0 \le A_V \le 4$ in equal steps of 0.02, and masses with a range
depending on the absolute magnitude of the source.  For this analysis,
we adopt the models of \citet{bc03} with a metallicity of
$0.4Z_{\odot}$.  We assume that the metallicity of the star clusters in the tidal 
debris matches the observed metallicity of tidal dwarf galaxies, $Z \sim 1/3
Z_{\odot}$ \citep{duc}, which is similar to the
metallicity of the outer regions of spiral galaxies.  
Table ~\ref{n2782ascph} lists the information for star cluster candidates
in the Eastern tail.  Tables ~\ref{n2782bsc} and ~\ref{n2782bscph} list the information for star cluster candidates
in the Western tail.  
Table \ref{n2782bsc} shows properties of the star clusters such as
reduced $\chi^2$, extinction, age, and mass, and their associated errors as well as the centroid pixel values. 
Table~\ref{n2782ascph} and \ref{n2782bscph} show star cluster photometry and errors 
in $UBVR$, colors, and $M_B$.

The errors in the magnitudes in each band were used to estimate the errors in the age, mass, and extinction.  For each $UBVR$ magnitude, we add the error in the magnitude to the magnitude and run the SED fitting code - the output ages, masses, and extinctions are the upper bound on the error bar for those values.  Then we subtract the error in the magnitude from each $UBVR$ magnitude and run the SED fitting code - the output ages, masses, and ext. are the lower bound on the error bar for those values.  For bright sources with small error bars, if the output age, mass, or extinction were the same as the data values, the upper and/or lower bounds for the errors were taken from the grid size of the models.  For example, a SCC may have an output age of 6.42 with the same age given when we run both ``error" models.  The next highest age bin is 6.44, so this is assigned to be the upper error and the next lowest age bin, 6.40, will be the lower error.  (See the star cluster candidates with their errors in age, mass, and extinction in Table \ref{n2782bsc}.) 

For the whole Eastern tail image field, out of 153 sources selected in
the above method, a total of 55 had reduced $\chi^2 < 3.0$ when fit to the
\citet{bc03} single stellar population models with 29 residing in the tail region.  
For the Western tail image field, 
out of 113 sources selected in the above method, a total of 49 had reduced
$\chi^2 < 3.0$ with 18 in the tail region.

In the ``out of tail" regions, 18 were fit in the Eastern tail image and
30 in the Western tail image.  These represent our contaminants.  Very
red sources fit with large extinction are likely to be foreground
stars that have similar colors.  Others may be background
galaxies.  To estimate the number of foreground stars in our images, 
we use the Besacon Milky Way star count model by \citet{robin}.  For location of NGC 2782 in the sky, the 6\arcmin \ field of view, and the range of colors and apparent magnitudes of our SCCs, we estimate there to be
about 13 foreground stars.  Since the Eastern tail contains about 40\% of the field of view, we then expect about 5
foreground stars within the tail and 8 in the ``out of tail" region.  The Western tail occupies about 30\% of the field of view, so we expect about 4 foreground stars in the ``in tail" region and 9 in the ``out of tail" region.  Using the Sloan Digital Sky Survey SkyServer DR6 Search \citep{sdssdr6} and the \citet{jester} transformations from $UBVR$ to $ugr$ filters, our range of color and magnitudes for the sources in the Eastern and Western tail estimate about 11 background galaxies in the Eastern tail entire field of view and about 18 in the Western tail field of view.   The transformation from $UBVR$ to $ugr$ filters is uncertain, so these numbers are only rough estimates for the background galaxies based on color and brightness.  In total, we estimate about 19 foreground stars and background galaxies for the Eastern tail (close to the 18 found in the ``out of tail" region) and 27 in the Western tail (close to the 30 found ``out of tail").  
The locations of the final star cluster candidates and the
``out of tail" sources fit by the SSP models are marked on the $V$ images
in Figures \ref{fig:VimageE} and \ref{fig:VimageW}.

%The level of contamination is ***.

%\clearpage
\begin{deluxetable}{ccccccccccc}
\rotate
\tablewidth{0pt}
\tabletypesize{\scriptsize}
\tablecaption{NGC 2782 East Star Cluster Candidate Photometry\label{n2782ascph}}
\tablehead{
  \colhead{\#}
  & X
  & Y
 & \colhead{$U$}
& \colhead{$B$}
& \colhead{$V$}
& \colhead{$R$}
& \colhead{$U-B$}
& \colhead{$B-V$}
& \colhead{$V-R$}
& \colhead{$M_B$}
}
\startdata
 
  0 &  394.152 &  565.963& 21.855 (0.069) &  22.343 (0.097) &  21.550 (0.087) &  21.027 (0.103) &  -0.488 (0.119) &  0.793 (0.130) &  0.522 (0.135) & -10.647 \\
  1 &  360.460 &  474.580 &  21.123 (0.039) &  20.998 (0.035) &  20.772 (0.038) &  20.394 (0.070) &   0.125 (0.052) &  0.226 (0.051) &  0.377 (0.080) & -11.992 \\
  2 &  398.524 &  507.196 &  21.269 (0.029) &  21.282 (0.028) &  21.109 (0.055) &  20.521 (0.064) &  -0.013 (0.041) &  0.172 (0.062) &  0.588 (0.084) & -11.708 \\
  3 &  429.452 &  456.452 &  22.053 (0.058) &  22.250 (0.068) &  21.736 (0.096) &  22.012 (0.265) &  -0.197 (0.090) &  0.514 (0.118) & -0.276 (0.281) & -10.740 \\
  4 &  383.717 &  448.772 &  22.285 (0.085) &  22.292 (0.092) &  23.056 (0.305) &  22.089 (0.303) &  -0.007 (0.125) & -0.764 (0.319) &  0.967 (0.430) & -10.698 \\
 61 &  196.373 &  192.751 &  23.320 (0.195) &  22.979 (0.115) &  22.529 (0.108) &  21.918 (0.116) &   0.341 (0.226) &  0.450 (0.158) &  0.611 (0.159) & -10.011 \\
 72&  285.484 &  236.534  &  23.175 (0.147) &  23.057 (0.072) &  22.724 (0.118) &  21.894 (0.122) &   0.118 (0.164) &  0.333 (0.138) &  0.830 (0.169) &  -9.933 \\
 86 &   95.453 &  288.121 &  23.745 (0.231) &  23.937 (0.163) &  22.978 (0.148) &  21.661 (0.100) &  -0.192 (0.282) &  0.959 (0.220) &  1.317 (0.179) &  -9.053 \\
146 &  380.222 &  357.056 &  23.619 (0.176) &  23.474 (0.132) &  22.652 (0.132) &  22.100 (0.176) &   0.146 (0.220) &  0.822 (0.187) &  0.552 (0.220) &  -9.516 \\
233 &  527.075 &  402.355 &  22.547 (0.091) &  22.002 (0.057) &  21.018 (0.063) &  20.286 (0.056) &   0.544 (0.108) &  0.984 (0.085) &  0.732 (0.084) & -10.988 \\
248 &  343.433 &  410.918 &  20.778 (0.022) &  20.426 (0.015) &  19.831 (0.017) &  19.445 (0.025) &   0.352 (0.027) &  0.595 (0.023) &  0.386 (0.031) & -12.564 \\
%317 &  23.233 (0.168) &  22.553 (0.109) &  21.734 (0.150) &  21.216 (0.196) &   0.680 (0.200) &  0.819 (0.186) &  0.518 (0.247) & -10.437 \\
358 &  367.965 &  467.203&  21.705 (0.081) &  21.590 (0.067) &  20.858 (0.048) &  20.410 (0.079) &   0.115 (0.104) &  0.732 (0.082) &  0.448 (0.092) & -11.400 \\
451 &  508.510 &  508.326&  23.259 (0.145) &  23.299 (0.117) &  22.577 (0.163) &  21.743 (0.184) &  -0.041 (0.186) &  0.723 (0.201) &  0.834 (0.246) &  -9.691 \\
506 &  512.762 &  536.482&  23.581 (0.214) &  23.424 (0.175) &  22.092 (0.144) &  21.492 (0.162) &   0.158 (0.277) &  1.332 (0.226) &  0.600 (0.217) &  -9.566 \\
530 &  367.374 &  549.503&  21.755 (0.044) &  22.137 (0.059) &  21.887 (0.125) &  21.191 (0.136) &  -0.382 (0.073) &  0.250 (0.138) &  0.696 (0.185) & -10.853 \\
543 &  399.654 &  556.366&  22.271 (0.091) &  22.348 (0.076) &  22.048 (0.129) &  21.258 (0.126) &  -0.077 (0.119) &  0.300 (0.149) &  0.790 (0.180) & -10.642 \\
555 &  370.081 &  562.678&  20.858 (0.028) &  20.996 (0.033) &  20.579 (0.045) &  20.358 (0.080) &  -0.138 (0.043) &  0.417 (0.056) &  0.222 (0.092) & -11.994 \\
596 &  408.717 &  595.590&  24.289 (0.389) &  23.949 (0.184) &  22.993 (0.187) &  21.622 (0.166) &   0.341 (0.430) &  0.955 (0.262) &  1.371 (0.250) &  -9.041 \\
611 &  349.476 &  607.032&  22.658 (0.078) &  22.272 (0.054) &  21.382 (0.055) &  20.347 (0.054) &   0.386 (0.095) &  0.891 (0.077) &  1.034 (0.077) & -10.718 \\
640 &  573.743 &  640.455 &  23.826 (0.218) &  23.435 (0.105) &  22.894 (0.133) &  22.026 (0.137) &   0.392 (0.242) &  0.541 (0.170) &  0.868 (0.191) &  -9.555 \\
650 &  221.230 &  646.889&  23.340 (0.177) &  22.784 (0.094) &  21.943 (0.113) &  21.266 (0.146) &   0.556 (0.201) &  0.842 (0.147) &  0.676 (0.185) & -10.206 \\
655 &  614.905 &  668.400&  23.104 (0.122) &  22.772 (0.054) &  21.566 (0.047) &  20.696 (0.040) &   0.332 (0.133) &  1.206 (0.071) &  0.870 (0.061) & -10.218 \\
665 &  101.749 &  702.029&  23.494 (0.158) &  23.433 (0.121) &  22.704 (0.123) &  21.983 (0.149) &   0.061 (0.199) &  0.729 (0.172) &  0.721 (0.193) &  -9.557 \\
682 &  127.326 &  743.661&  24.429 (0.472) &  23.557 (0.159) &  22.433 (0.121) &  21.004 (0.071) &   0.872 (0.498) &  1.124 (0.200) &  1.428 (0.140) &  -9.433 \\
684 &  619.316 &  749.198&  23.525 (0.184) &  23.915 (0.150) &  22.963 (0.172) &  22.324 (0.169) &  -0.390 (0.238) &  0.952 (0.228) &  0.639 (0.241) &  -9.075 \\
690 &  513.713 &  761.758&  24.085 (0.285) &  23.810 (0.154) &  22.902 (0.128) &  22.095 (0.126) &   0.275 (0.324) &  0.907 (0.200) &  0.807 (0.179) &  -9.180 \\
695 &  362.009 &  777.885 &  22.894 (0.099) &  22.835 (0.059) &  22.276 (0.075) &  22.125 (0.125) &   0.060 (0.115) &  0.558 (0.095) &  0.151 (0.145) & -10.155 \\
  
 \enddata
 \end{deluxetable}

%%\clearpage
\begin{deluxetable}{ccccccc}
\tabletypesize{\scriptsize}
\tablewidth{0pt}
\tablecaption{NGC 2782 West Star Cluster Candidate Properties\label{n2782bsc}}
\tablehead{
  \colhead{Number}
& \colhead{$\chi ^2$}
& \colhead{Extinction}
& \colhead{Age} 
& \colhead{Mass}
& \colhead{$X$}
& \colhead{$Y$}
\\

  \colhead{}
& \colhead{}
& \colhead{$E(B-V)$}
& \colhead{log(yr)} 
& \colhead{log($M/M_\odot$)}
& \multicolumn{2}{c}{pixel}
%& \multicolhead{15}{c}{mag.}

}
\startdata

 97 &  0.1 &  0.50$^{ 0.49}$$_{ 0.51}$ &  7.58$^{ 7.54}$$_{ 7.63}$ &  5.67$^{ 5.68}$$_{ 5.66}$ &  501.704 &  266.439 \\\\
101 &  0.6 &  0.27$^{ 0.26}$$_{ 0.21}$ &  8.06$^{ 8.01}$$_{ 8.21}$ &  5.73$^{ 5.75}$$_{ 5.70}$ &  498.146 &  270.219 \\\\
120 &  0.0 &  0.63$^{ 0.61}$$_{ 0.64}$ &  7.76$^{ 7.70}$$_{ 7.81}$ &  6.15$^{ 6.10}$$_{ 6.16}$ &  421.899 &  338.391 \\\\
152 &  2.5 &  0.97$^{ 0.96}$$_{ 0.98}$ &  6.40$^{ 6.38}$$_{ 6.42}$ &  5.96$^{ 5.97}$$_{ 5.95}$ &  334.414 &  421.170 \\\\
%154 &  1.1 &  1.03$^{ 1.01}$$_{ 0.74}$ &  6.26$^{ 6.24}$$_{ 7.00}$ &  6.16$^{ 6.15}$$_{ 5.75}$ &  334.414 &  421.170 \\\\
172 &  1.3 &  0.00$^{ 0.00}$$_{ 0.01}$ &  8.46$^{ 8.51}$$_{ 8.40}$ &  5.74$^{ 5.82}$$_{ 5.70}$ &  451.681 &  477.299 \\\\
182 &  0.2 &  0.99$^{ 0.98}$$_{ 1.01}$ &  6.44$^{ 6.40}$$_{ 6.46}$ &  5.75$^{ 5.82}$$_{ 5.74}$ &  541.170 &  494.540 \\\\
185 &  0.7 &  1.21$^{ 1.18}$$_{ 1.25}$ &  6.34$^{ 6.32}$$_{ 6.36}$ &  6.06$^{ 6.07}$$_{ 6.05}$ &  489.818 &  501.482 \\\\
204 &  0.0 &  1.42$^{ 1.38}$$_{ 1.46}$ &  6.34$^{ 6.36}$$_{ 6.26}$ &  6.57$^{ 6.55}$$_{ 6.65}$ &  348.104 &  552.657 \\\\
226 &  0.5 &  0.01$^{ 0.03}$$_{ 0.00}$ &  7.53$^{ 7.49}$$_{ 7.54}$ &  5.05$^{ 5.07}$$_{ 5.01}$ &  536.171 &  599.457 \\\\
227 &  0.9 &  0.26$^{ 0.82}$$_{ 0.28}$ &  9.30$^{ 7.86}$$_{ 9.32}$ &  7.06$^{ 6.70}$$_{ 7.08}$ &  764.120 &  600.468 \\\\
234 &  0.0 &  0.94$^{ 0.92}$$_{ 0.96}$ &  6.40$^{ 6.42}$$_{ 6.32}$ &  5.66$^{ 5.67}$$_{ 5.76}$ &  810.621 &  623.717 \\\\
235 &  0.6 &  0.57$^{ 0.58}$$_{ 0.64}$ &  6.58$^{ 6.60}$$_{ 6.22}$ &  5.35$^{ 5.39}$$_{ 5.80}$ &  545.504 &  626.576 \\\\
282 &  0.0 &  0.17$^{ 0.20}$$_{ 0.15}$ &  7.86$^{ 7.76}$$_{ 7.91}$ &  5.30$^{ 5.34}$$_{ 5.25}$ &  584.557 &  710.380 \\\\
299 &  0.1 &  0.78$^{ 0.77}$$_{ 0.75}$ &  7.91$^{ 7.81}$$_{ 8.11}$ &  6.56$^{ 6.52}$$_{ 6.60}$ &  558.372 &  730.295 \\\\
331 &  1.2 &  0.82$^{ 0.81}$$_{ 0.83}$ &  6.40$^{ 6.44}$$_{ 6.38}$ &  5.48$^{ 5.46}$$_{ 5.45}$ &  452.459 &  767.836 \\\\
398 &  2.4 &  0.57$^{ 0.55}$$_{ 0.52}$ &  7.86$^{ 7.72}$$_{ 8.11}$ &  5.94$^{ 5.88}$$_{ 5.97}$ &  548.341 &  836.734 \\\\
412 &  0.2 &  1.21$^{ 1.22}$$_{ 1.20}$ &  7.32$^{ 7.26}$$_{ 7.42}$ &  6.98$^{ 6.95}$$_{ 7.01}$ &  641.640 &  850.058 \\\\
448 &  0.6 &  0.75$^{ 0.71}$$_{ 0.81}$ &  6.82$^{ 6.84}$$_{ 6.80}$ &  5.39$^{ 5.37}$$_{ 5.41}$ &  749.070 &  880.521 \\\\
%448 &  0.6 &  0.75$^{ 0.71}$$_{ 0.81}$ &  6.82$^{ 6.84}$$_{ 6.80}$ &  5.39$^{ 5.37}$$_{ 5.41}$ &  749.070 &  880.521 \\\\
  
 \enddata
 \end{deluxetable}
%%\clearpage
\ \begin{deluxetable}{cccccccccccccccc}
\rotate
\tablewidth{0pt}
\tabletypesize{\scriptsize}
\tablecaption{NGC 2782 West Star Cluster Candidate Photometry\label{n2782bscph}}
\tablehead{
  \colhead{\#}
& \colhead{$U$}
& \colhead{$B$}
& \colhead{$V$}
& \colhead{$R$}
& \colhead{$U-B$}
& \colhead{$B-V$}
& \colhead{$V-R$}
& \colhead{$M_B$}
%\\

%  \colhead{}
}
\startdata

97 &  23.274 (0.134) &  23.411 (0.099) &  22.814 (0.092) &  22.290 (0.079) &  -0.136 (0.166) &  0.596 (0.135) &  0.524 (0.121) &  -9.579 \\
101 &  22.967 (0.102) &  23.061 (0.074) &  22.610 (0.081) &  22.327 (0.078) &  -0.094 (0.126) &  0.451 (0.109) &  0.283 (0.112) &  -9.929 \\
120 &  23.076 (0.111) &  23.016 (0.065) &  22.330 (0.072) &  21.690 (0.051) &   0.060 (0.129) &  0.686 (0.097) &  0.640 (0.088) &  -9.974 \\
152 &  22.503 (0.080) &  22.978 (0.065) &  22.375 (0.067) &  21.569 (0.046) &  -0.475 (0.103) &  0.603 (0.093) &  0.806 (0.081) & -10.012 \\
172 &  22.635 (0.085) &  22.573 (0.061) &  22.369 (0.104) &  22.307 (0.140) &   0.062 (0.105) &  0.204 (0.120) &  0.063 (0.174) & -10.417 \\
182 &  23.058 (0.120) &  23.488 (0.108) &  22.739 (0.098) &  22.066 (0.075) &  -0.430 (0.161) &  0.749 (0.145) &  0.673 (0.123) &  -9.502 \\
185 &  23.621 (0.168) &  23.832 (0.153) &  23.045 (0.120) &  22.063 (0.082) &  -0.211 (0.227) &  0.787 (0.195) &  0.983 (0.145) &  -9.158 \\
204 &  23.398 (0.155) &  23.429 (0.100) &  22.347 (0.082) &  21.309 (0.039) &  -0.031 (0.184) &  1.082 (0.129) &  1.037 (0.091) &  -9.561 \\
226 &  22.376 (0.063) &  22.841 (0.062) &  22.883 (0.113) &  22.599 (0.108) &  -0.465 (0.088) & -0.042 (0.129) &  0.283 (0.156) & -10.149 \\
227 &  22.917 (0.134) &  22.637 (0.042) &  21.658 (0.040) &  20.944 (0.025) &   0.281 (0.140) &  0.979 (0.058) &  0.714 (0.047) & -10.353 \\
234 &  23.163 (0.175) &  23.585 (0.108) &  22.918 (0.135) &  22.297 (0.095) &  -0.422 (0.206) &  0.666 (0.173) &  0.621 (0.165) &  -9.405 \\
235 &  21.903 (0.040) &  22.492 (0.047) &  22.075 (0.059) &  21.741 (0.055) &  -0.589 (0.062) &  0.417 (0.075) &  0.334 (0.080) & -10.498 \\
282 &  23.171 (0.129) &  23.404 (0.102) &  23.143 (0.148) &  22.888 (0.141) &  -0.234 (0.165) &  0.261 (0.180) &  0.255 (0.204) &  -9.586 \\
299 &  23.064 (0.142) &  22.793 (0.075) &  21.980 (0.083) &  21.192 (0.050) &   0.270 (0.161) &  0.814 (0.112) &  0.788 (0.097) & -10.197 \\
331 &  22.991 (0.112) &  23.660 (0.128) &  22.999 (0.113) &  22.461 (0.095) &  -0.669 (0.170) &  0.661 (0.171) &  0.538 (0.148) &  -9.330 \\
398 &  23.519 (0.184) &  23.348 (0.109) &  22.990 (0.136) &  22.152 (0.085) &   0.171 (0.214) &  0.358 (0.174) &  0.838 (0.161) &  -9.642 \\
412 &  22.849 (0.099) &  22.556 (0.046) &  21.377 (0.028) &  20.245 (0.015) &   0.293 (0.109) &  1.179 (0.054) &  1.132 (0.032) & -10.434 \\
448 &  23.296 (0.173) &  23.358 (0.088) &  22.350 (0.063) &  21.599 (0.046) &  -0.062 (0.194) &  1.008 (0.109) &  0.751 (0.078) &  -9.632 \\
 
 \enddata
 \end{deluxetable}
\clearpage

\subsection{{\Ha} Images}
Narrow-band {\Ha} images were also obtained at the \emph{VATT} with an 88 mm
Andover 3-cavity interference filter centered at 6630 \AA.  The FWHM of the filter is $70\AA$ which includes the [NII] lines.
Integration times for the {\Ha} images were $3 \times 1200$ s on the
Eastern tail and $2 \times 1200$ s and $6\times 900$ on the Western tail.  To subtract
continuum emission, both fields were also observed with a Kron-Cousins
$R$ filter using integration times of $3 \times 300$ s.  Images were
reduced in a similar manner as stated above.

To create images with only the emission lines, a scaled $R$ band image
was subtracted from narrowband image after alignment using foreground
stars.  To determine an initial scaling factor, the ratio of
integration time for individual frames is multiplied by the ratio of
filter widths.  If there were still strong continuum features in the
galaxies, the scaling factor was varied iteratively until the
continuum-dominated regions in the galaxy and the background source
and foreground star residuals reached a minimum \citep{lee}. The insets in Figures~\ref{fig:VimageE} and \ref{fig:VimageW} show the continuum-subtracted {\Ha}  images for the regions of interest.

\subsubsection{Calibration of {\Ha}}
Spectrophotometric standard stars from the \citet{okestone} catalog
were observed on each night.  Aperture photometry of these standards
was compared to their absolute magnitudes.  Absolute magnitudes for
each spectrophotometric standard star were calculated by integrating
their spectral energy distribution over the filter response
function. As in \citet{lee}, we used a standard atmospheric extinction
coefficient of 0.08 mag airmass$^{-1}$.  Zeropoints were calculated by
comparing the absolute magnitude in each filter
with the instrumental magnitude from aperture photometry.  For each
night, the zeropoints from all standards (typically 3-5) were
averaged.  

Due to the proximity of the [NII] doublet at $\lambda$6548,6583 to the
{\Ha} line, their contribution to the flux needs to be accounted for and
removed.  Also, the use of the broadband $R$ filter for the continuum
measurement means that emission line flux from that filter needs to be
removed.  The total flux equation (Equation A13 in \citet{lee}) is:

\begin{eqnarray}
\lefteqn{f_{tot}(\rm H\alpha + [NII]) = } \nonumber \\
& \lambda^{-2}10^{-0.4(ZP+2.397-\kappa sec(z))} FWHM_{NB} CR(\rm H\alpha+[NII]) \left[ T_{NB}(\lambda) - T_R(\lambda)
\frac{t_R}{t_{NB}}\frac{1}{F}\right] ^{-1} 
\end{eqnarray}

\noindent $\lambda$: redshifted wavelength of {\Ha} \\
 $ZP$: zero point \\
 $\kappa$: atmospheric extinction coefficient (we use 0.08 mag
airmass$^{-1}$) \\
 $FWHM_{NB}$: width of narrowband filter in \AA \\
 $CR(\rm H\alpha+[NII])$: count rate in continuum subtracted image \\
 $T_R$: Transmission correction in $R$.  Calculated by an average of
normalized transmissions at each redshifted wavelength of {\Ha} and
[NII] lines, weighted by their relative line fluxes.   \\
 $T_{NB}$: Transmission correction for narrowband filter.  Calculated
by an average of normalized transmissions at each redshifted
wavelength of {\Ha} and [NII] lines, weighted by their relative line
fluxes. \\
 $t_R$: exposure time in $R$ band filter \\
 $t_{NB}$: exposure time in  narrowband filter  \\
 $F$: scale factor applied to $R$ band continuum image when subtracting
it from narrowband image.

While $T_{NB}$ would ideally be calculated by measuring the line
ratios of {\Ha} and [NII] lines directly from spectroscopy, in the
absence of spectra for HII regions, there is a relation between
metallicity and the ratio of [NII] to {\Ha}.  In Figure 9 of
\citet{vanZee98}, an empirical relation is shown between these two values:  

\begin{equation}
12 + \rm log (O/H) = 1.02 \rm log ([NII]/H\alpha) + 9.36
\end{equation}
%justify
For the metallicity of $0.4Z_{\odot}$, as used in the single stellar
population models to fit the star cluster candidates, $12 + \rm log (O/H)
= 8.06$ which gives $\rm log ([NII]/H\alpha) = -1.3$.  These numbers were
also used to subtract the flux of the [NII] lines, giving a resulting
flux that contains only that of {\Ha}.  If we instead use the value of $12 + \rm log (O/H)
= 8.72$ as measured by \citet{werk} in the Western tail of NGC 2782, we find 
$\rm log ([NII]/H\alpha) = -0.63$.  Using this value to determine our {\Ha} flux, 
we find that for the Western tail region the difference between the {\Ha} luminosity 
for the higher metallicity ratio and for the  lower metallicity ratio is 
$0.1\times 10^{38}$ erg {\persec} which is less than the error bar ($\pm0.4\times 10^{38}$ erg \persec).
We adopt the lower metallicity value for this work.

\subsection{{\CII} observations}
To map the {\CII} 158$\mu$m fine structure line in the tidal tails of NGC 2782, we used the PACS spectrometer \citep{poglitsch}
on the \emph{Herschel Space Observatory} \citep{pilbratt}.   The {\CII} 158$\mu$m line is observed with the 2nd plus 
1st order gratings.  We used
pointed observations with chopping/nodding and a large throw of 6{\arcmin} off the source.  To increase the line sensitivity, we set 
the line repetition factor to the maximum value of 10 for each target.  To increase the total sensitivity, we repeat each 
observation cycle 3 times.  To reach the needed depth, we observed each pointing for 3.1 hours.  

We observed at one location in each tidal tail corresponding to areas of recent star formation as shown by {\Ha} emission indicated in our narrowband images.  The PACS array has a total field of view of $47{\arcsec} \times 47{\arcsec}$ with 5 $\times$ 5 spatial pixels of 9.4{\arcsec} size.  Each spatial pixel has 16 spectral elements.  This size easily encompasses the star forming region in the Western tail of NGC 2782 
\citep{knierman12}.  The Eastern tail of NGC 2782 has a larger spatial extent of star formation; we choose to 
target only the area with the highest observed CO and HI.  

\begin{figure}
\includegraphics[width=\textwidth]{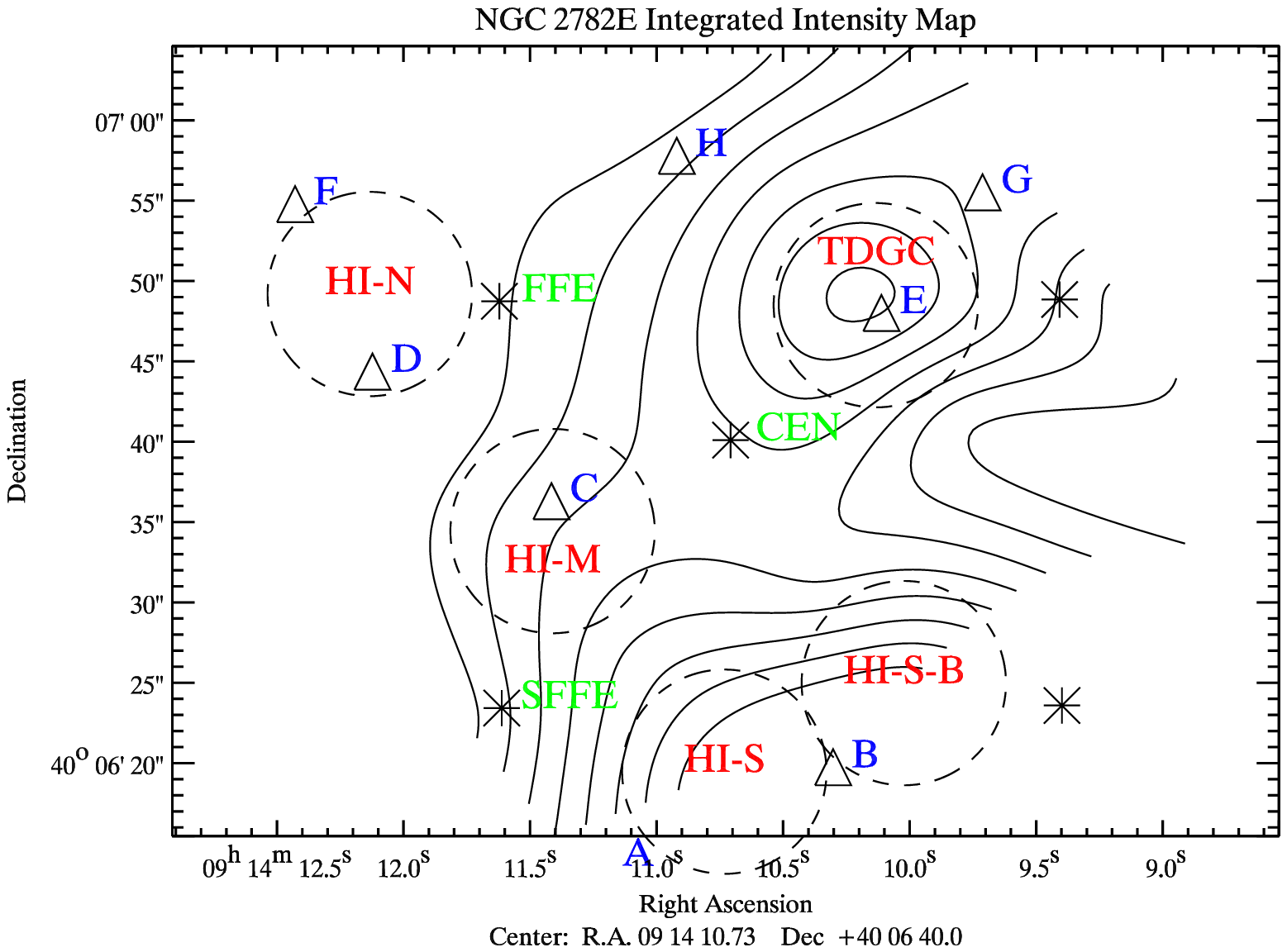}
\caption{Contour of {\CII} in Eastern tail of NGC 2782 from Herschel/PACS.  Solid contours range from $3-10\sigma$ with $\sigma = 0.0028$ Jy $\mu$m pix$^{-1}$.  Circles with red labels indicate position of extracted spectra shown in Fig. \ref{fig:spectra}. Triangles with blue labels mark the locations of {\Ha} sources found here and also listed in \citet{smith99}.  Asterisks mark the locations where CO(1-0) was looked for with the Kitt Peak 12 meter telescope with the green labeled points indicating locations discussed in the text.  The location labeled ``CEN" was observed in this work (Fig.~\ref{fig:CO}) while all other locations were observed by \citet{smith99}. }
\label{fig:contour}
\end{figure}

\begin{figure}
\includegraphics[width=\textwidth]{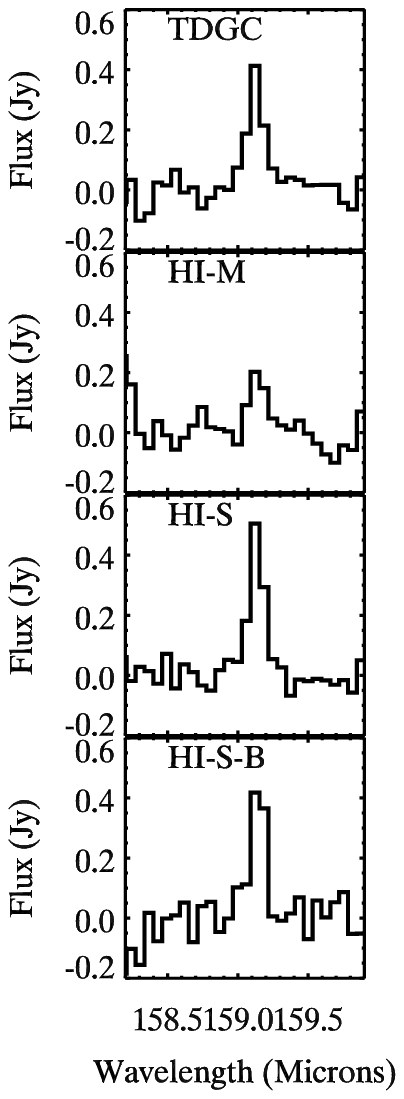}
\caption{Spectra of {\CII} extracted at locations in Eastern tail indicated in Fig. \ref{fig:contour}.}
\label{fig:spectra}
\end{figure}

The data were processed using the PACS spectrometer pipeline of the Herschel Interactive Processing Environment (HIPE) Version 7.3.0.  The resulting PACS Rebinned Spectral Cube product was exported to a fits cube for further analysis using IDL routines.  The {\CII} line emission peaks at 159.123 $\mu$m for the sources in the Eastern tail.   After subtracting a linear baseline, we produced an integrated intensity map (Fig.~\ref{fig:contour}) by integrating over the wavelength range from  158.997-159.249$\mu$m.  Contours from 3-10$\sigma$ are plotted with $\sigma=0.0028$ Jy $\mu$m pix$^{-1}$.  We determine $\sigma$ by taking the rms of spectra outside the line area, multiplying by the square root of the number of channels integrated for the map (5 in this case), and the width of one channel, 0.063 $\mu$m.  
Spectra (as shown in Fig.~\ref{fig:spectra}) were extracted from a single spaxel from the PACS Rebinned Spectral Cube at the locations indicated in Fig. ~\ref{fig:contour}.  After subtracting a linear baseline, the line intensities were summed over the same range as above and multiplied by the bandwidth.  Errors in the intensity were calculated from the rms value outside the line area.   To correct the amount of flux that falls outside the single spaxel, we use a flux correction of $1/0.5$ at 160 $\mu$m according to Fig. 7 of the PACS Spectroscopy Performance and Calibration Version 2.4 document \citep{pacscalib}.  This aperture correction assumes that the source is a point source lying at the center of the given spaxel.

\subsection{CO(1-0) observations}
\begin{figure}
\includegraphics[width=\textwidth]{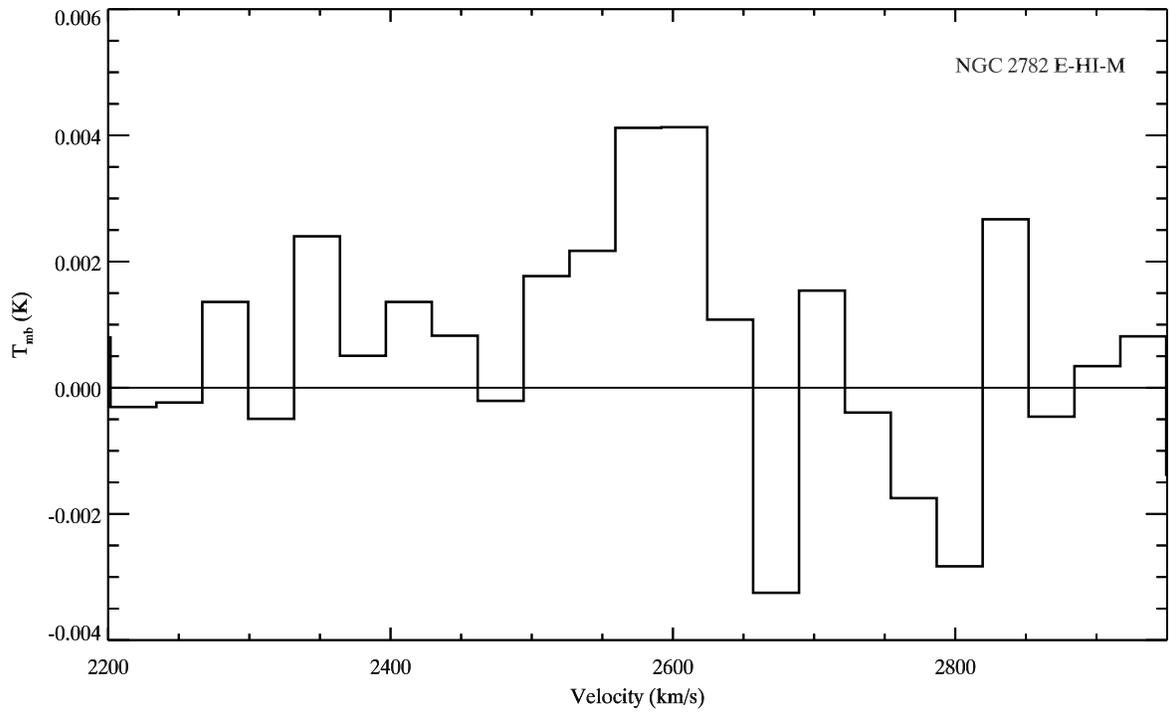}
\caption{Spectra of CO(1-0) in Eastern tail of NGC 2782 from ARO Kitt Peak 12 meter telescope. The location of the observation is the same as the {\CII} observation with  Herschel.}
\label{fig:CO}
\end{figure}

The Eastern tail and Western tail of NGC 2782 were observed in the CO(1-0) line in March and October 2012 
using the Arizona Radio Observatory Kitt Peak 12 meter telescope.  We used the ALMA band 3 receiver with 2IFs (both 
polarizations in USB) using
the MAC autocorrelator in 800MHz bandwidth mode with 2048 channels.  The total bandpass is
1500 km s$^{-1}$ wide and was centered at 2555 km s$^{-1}$.  We observed in position switch mode with a 10{\arcmin} throw in azimuth to the off position.   
%with a spectral resolution of **.  
The beam size FWHM is 55{\arcsec} at 115 GHz.  The system temperature ranged from 270-360K.  % (**)??
We observed at the same locations of recent star formation in the Eastern and Western tails of NGC 2782 as our [CII] observations.  For both tails, we reach an rms value of 1 mK which is deeper than the observations of these tails by \citet{smith99}.  The final summed scan for the Eastern tail location is shown in Fig. \ref{fig:CO}.  

\section{Results}
As seen in previous work \citep{smith94,wehnerthesis}, the deep optical
images presented here show that NGC 2782 has an Eastern tail
stretching 20 kpc to the northeast.  This tail consists of a bright
region at the tip with a fainter, yet clumpy bridge region between the
main body of the spiral and the bright region at the tip of the tail.
There is also a fainter extension to the southeast as well as faint
debris to the north and south of the main body of the spiral.  The
Western tail stretches 50 kpc to the northwest.  It has a faint yet
clumpy structure along its length.  There is a single stream
stretching the full 50 kpc as well as a smaller extension to the north
which only extends one-third as far.

We now examine several different tracers of star formation between the tidal tails: the young star cluster population, {\HII} regions, and the gas properties including neutral, ionized, and molecular gas.

\subsection{Properties of star clusters in the tidal debris of NGC 2782}
The final sample of star cluster candidates include 28 sources in the
Eastern tail and 19 sources in the Western tail.  As seen in Figure
\ref{fig:VimageE}, the majority (70\%) of the star cluster candidates
found in the Eastern tail are found in the bridge region.  Four
candidates are found in the southeast extension while 3 are found in
the debris to the north and 3 are found south of the central galaxy.
In the Western tail (Fig. \ref{fig:VimageW}), the star cluster candidates
are mostly located along the main tail with only 2 in the short tail
to the north.

\subsubsection{Overdensity of star clusters in the tails}
While it is difficult to remove further contaminants to the star
cluster sample without using additional observations (e.g.,
spectroscopy), an idea of the amount of star clusters in the tidal
tails can be determined statistically by subtracting the background
density from the ``in tail" sample.  This technique was used in
\citet{knierman} to determine the overdensity of star clusters in the
tidal tails of major mergers.  In that paper, young star clusters were
selected by using $M_V<-8.5$ and $V-I<0.7$ since only one color was
available.  For this study, we again use $M_V<-8.5$, but since we are
unable to match the color cut exactly, we take the \citet{bc03} model
age of log(age)=8.5 at which $V-I=0.7$ as our criteria.  Table~\ref{backsub2782} gives
the calculation of the overdensity of young star clusters in the tidal
debris.  Both the Eastern and Western tails have similar values for
the overdensity of 0.005-0.006 star clusters per kpc$^2$.  This is much
lower than the value of 0.108 kpc$^{-2}$ for the Western tail of major merger NGC
3256.  Given the differing areas in the Eastern and Western tails, the
expected number of actual star clusters is 14 and 10 respectively.
In their HST/WFPC2 survey \citet{mullan}, also calculate the overdensity of these tails, but for 
smaller regions within one WFPC2 pointing.  Their values are 0.234 and 0.016 kpc$^{-2}$, respectively.
These values are 39 and 3 times larger than the values measured here.  We suspect the difference in areas 
may have a large affect on this calculation so we calculate the overdensity of our star cluster candidates within 
area of their WFPC2 observation.  We find 9 star cluster candidates within this region of the tail, and only 1 that overlaps in the ``out of tail" region.  This gives an overdensity of $0.02\pm 0.01$ for the Eastern tail, now only a factor of ten lower than the value from \citet{mullan}.  

%more here about Brendan's work, but his numbers are higher than mine...??  area?

%\clearpage
\begin{deluxetable}{ccccccccccc}
\tabletypesize{\scriptsize}
\tablewidth{6in}
\tablewidth{0pt}
\tablecaption{Overdensity of star clusters in tails}
%\vglue -0.5in
%\tablenum{4}

\tablehead{\colhead{Tail} & \colhead{Pixel size} & \colhead{$N_{in}$} & \colhead{Area$_{in}$} & \colhead{Area$_{in}$} & \colhead{$N_{in}$/kpc$^{2}$} & \colhead{$N_{out}$} & \colhead{Area$_{out}$} & \colhead{Area$_{out}$} & \colhead{$N_{out}$/kpc$^{2}$} & \colhead{Surplus} \\
\colhead{} & \colhead{pc} & \colhead{}& \colhead{pix$^{2}$} & \colhead{kpc$^{2}$} & \colhead{} & \colhead{} &  \colhead{pix$^{2}$} & \colhead{kpc$^{2}$} & \colhead{} & \colhead{} }
\startdata
E & 80.69 & 20 & 345877 & 2252.0 & 0.009(0.002) & 9 & 475199 & 3094.0 & 0.003(0.001) & 0.006(0.002)\\
W & 80.69 & 17 & 257238 & 1674.9 & 0.010(0.002) & 19 & 584730 & 3807.2 & 0.005(0.001) & 0.005(0.003)
\enddata

\label{backsub2782}

\end{deluxetable}

%\subsection{Comparison of the two tails and their clusters}

%
\subsection{Properties of Star Cluster Candidates}
\begin{figure}
\includegraphics[width=\textwidth]{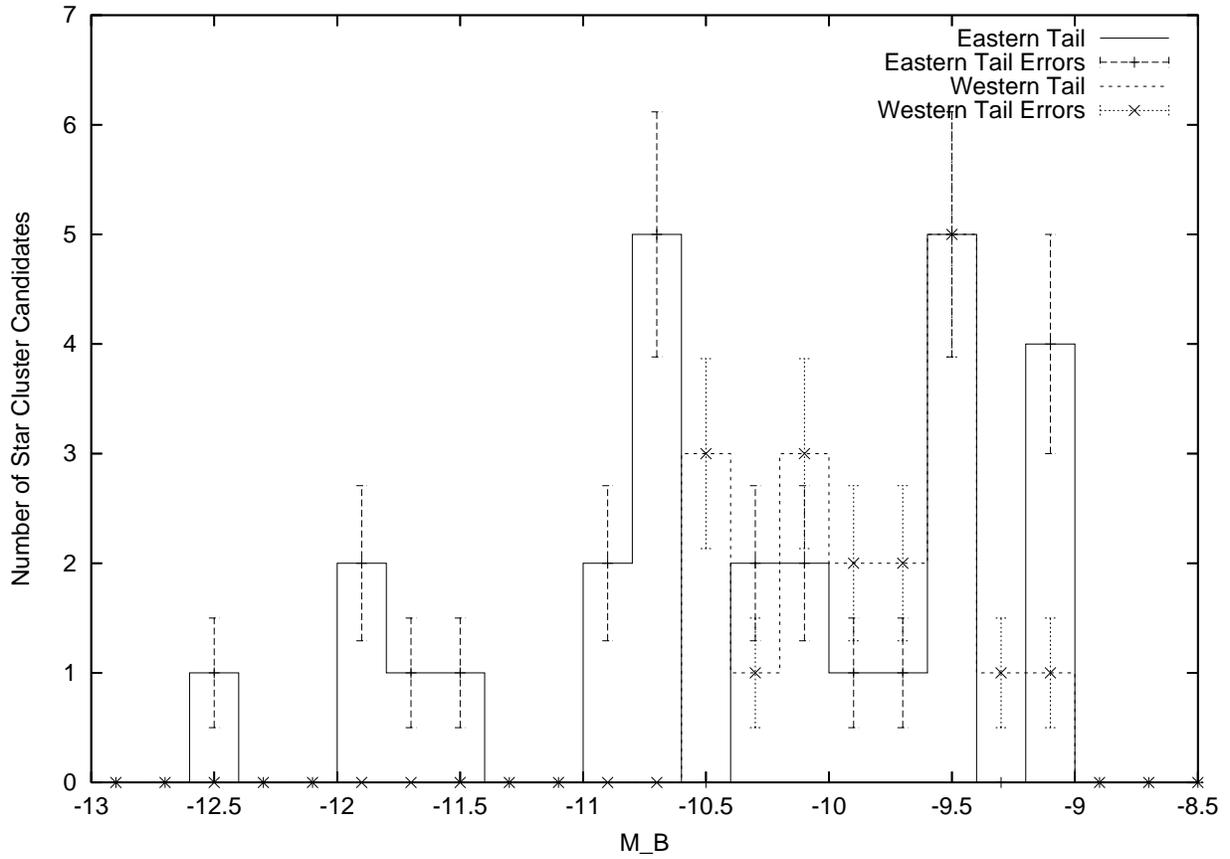}
\caption{Histogram of $M_B$ for final star cluster candidates.  
The solid line indicates Eastern tail candidates
and the dotted line indicates Western tail candidates.  Error bars represent the Poisson error for each bin:  the square root of the number of star cluster candidates in the bin.}
\label{fig:hist_Blum}
\end{figure}

As shown in Figure~\ref{fig:hist_Blum}, more luminous star cluster
candidates are found in the Eastern tail than in the Western tail.
The ranges of luminosity are $-12.6 < M_B < -9$ for the Eastern tail
and $-10.5 < M_B < -9$ for the Western tail.  The average $M_B$ for
the Eastern tail is -10.34, while the average $M_B$ in the Western
tail is -9.85.  According to the Kolmogorov-Smirnov test, these
distributions of $M_B$ have a probability of $P=0.019$ of being drawn
from the same distribution.  

%The ranges for the luminosities of the star cluster candidates is within the
%luminosity ranges of $-12 < M_B < -9$ for ``super star clusters''
%\citep{ferreiro, duc04}. 

\citet{whitmore99} use a faint end cutoff of
$M=-9$ since sources fainter than that limit are likely to be bright
stars at the distance of the galaxy.  
The brighter end ($M_B = -12$) corresponds to
the luminosity of {\HII} regions in Sa/Sb galaxies \citep{bk97} and to an age of
$\sim60$ Myr in the \citet{bc03} models for a $10^6$ {\msun} star
cluster.
These star cluster candidates are all much more luminous than the Trapezium
cluster in Orion which has $M_B=-4$.  The $B$ magnitude for 30 Dor in
the Large Magellanic Cloud is $B= 9.63$ which at a distance of 50 kpc
gives $M_B = -8.86$.  So these star clusters are also more luminous
than 30 Dor.

Since the Eastern tail has star cluster candidates with brightnesses similar to {\HII} regions
and our ground based observations may hide the extended nature of sources in the tidal tails,
we examine the HST/WFPC2 images of both tails from \citet{mullan}.  As shown in their Fig. 3.7 and Fig. 3.8,
the brighter Eastern tail sources are extended with multiple bright clumps, but the Western tail sources remain compact and isolated.  Therefore, our ground based observations of the Western tail sources are able to be treated as single star cluster candidate with one stellar population, particularly since the HST observations are only in F606W and F814W bands.  However, we must treat the Eastern tail sources differently due to their extended nature. 

\begin{figure}
\plottwo{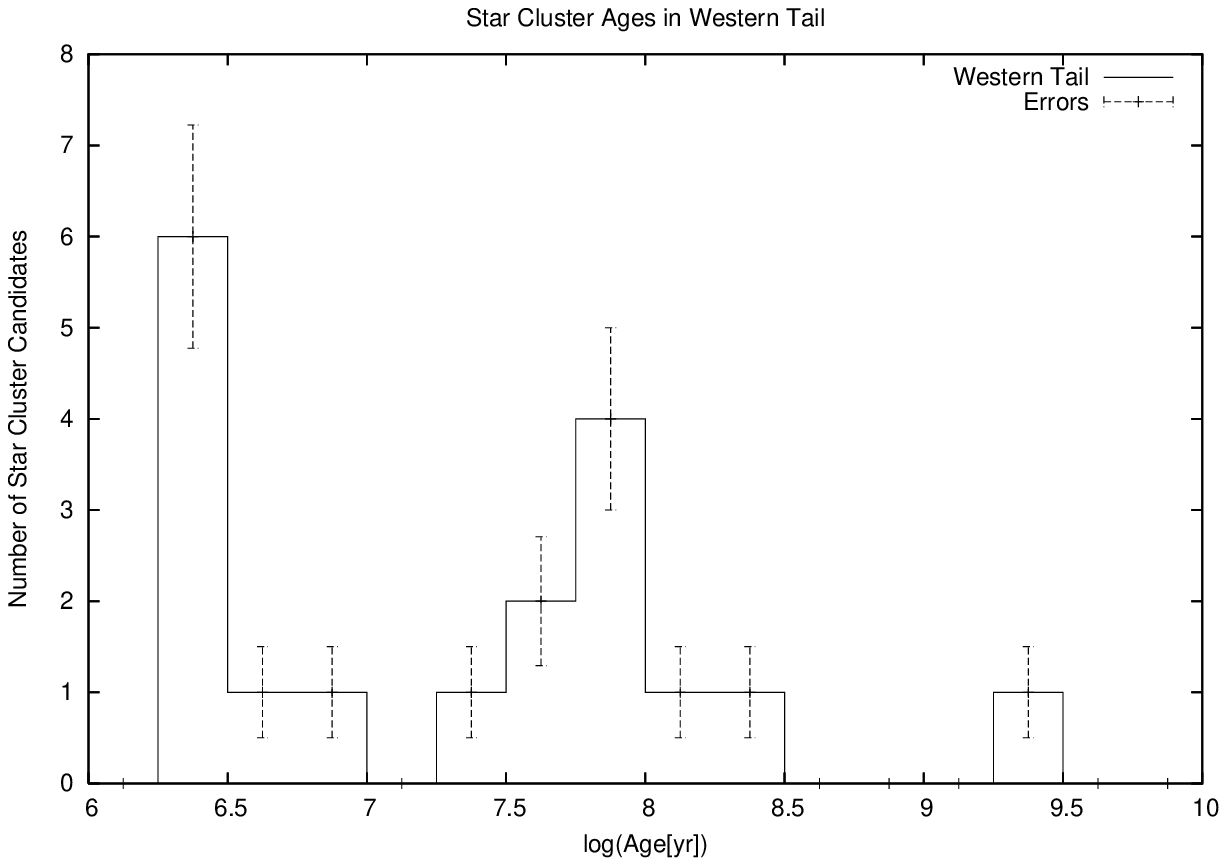}{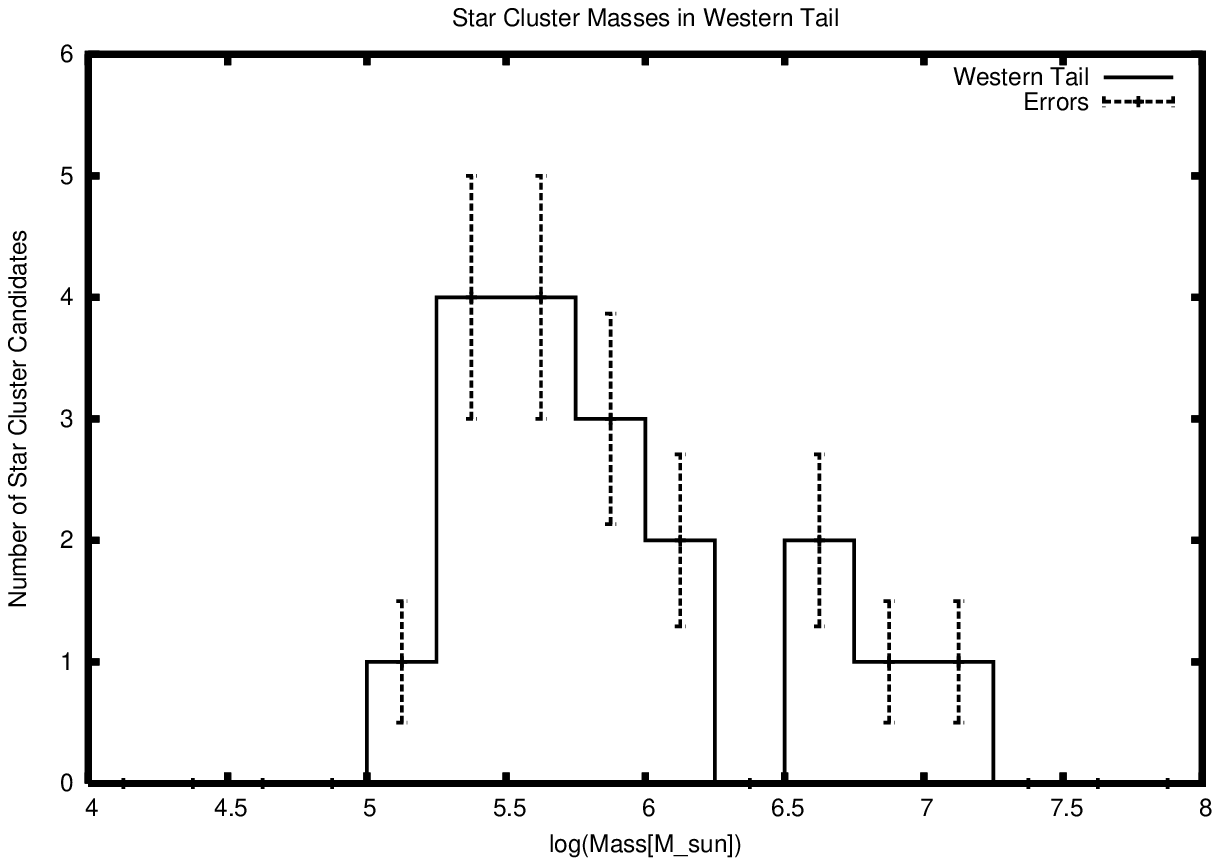}
\caption{Left: Histogram of star cluster candidate ages for the Western tail of NGC 2782.  Right: Histogram of star cluster candidate masses for the Western tail of NGC 2782.  Error bars represent the Poisson error for each bin:  the square root of the number of star cluster candidates in the bin.}
\label{fig:hist_age_mass}
\end{figure}

Due to the compact and isolated nature of the Western tail sources, we continue with the 3DEF SED fitting method to determine ages, masses, and extinction for this tail.  
As shown in Figure~\ref{fig:hist_age_mass}, star cluster candidates in the Western tail
tails have ages ranging from 2.5 Myr to 1 Gyr.  The median age in the Western
tail is 150 Myr.  
This tail has a significant fraction (90\%) of  star cluster candidates whose ages are less than the
age of the merger $\sim200$ Myr (log(age[yr]) = 8.3).  Therefore, there are star clusters that formed in situ in
the Western tidal tail of NGC 2782.  Figure~\ref{fig:hist_age_mass} shows the range of star cluster
candidate masses in the Western tail.  The Western tail has star
cluster candidates ranging in mass from $10^5-10^{7.25}$ \msun, with a
median mass of $5\times10^5$ \msun.  The masses of the star cluster candidates in this tail
are much larger than the $10^3$ {\msun} of the Orion cluster and the
$2\times10^4$ {\msun} mass of R136, but are on the order of masses of
star clusters in central starburst of the Antennae galaxy.

\subsection{Star Cluster Complexes in Eastern Tail}
Stars often form in pairs or clusters, not in isolation, and star clusters may also form in groups.
These groups, called star cluster complexes, represent one of the levels of hierarchical star formation in galaxies
and are often seen in spiral galaxies. 
In nearby spiral galaxies, \citet{elmsalzer} find a linear relation between the brightness of a star cluster complex and its linear size.  
In optical and molecular observations of M51, \citet{bastianm51} find that star cluster complexes are young ($<10$ Myr) and have a strong correlation between mass and size (unlike for isolated star clusters) which is similar to that found for GMCs.   In studies of Hickson Compact Group galaxies, \citet{isk59} find a linear relation in HCG 59, but no relation for HCG 7 \citep{isk7}.

\begin{figure}
\plottwo{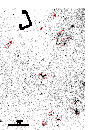}{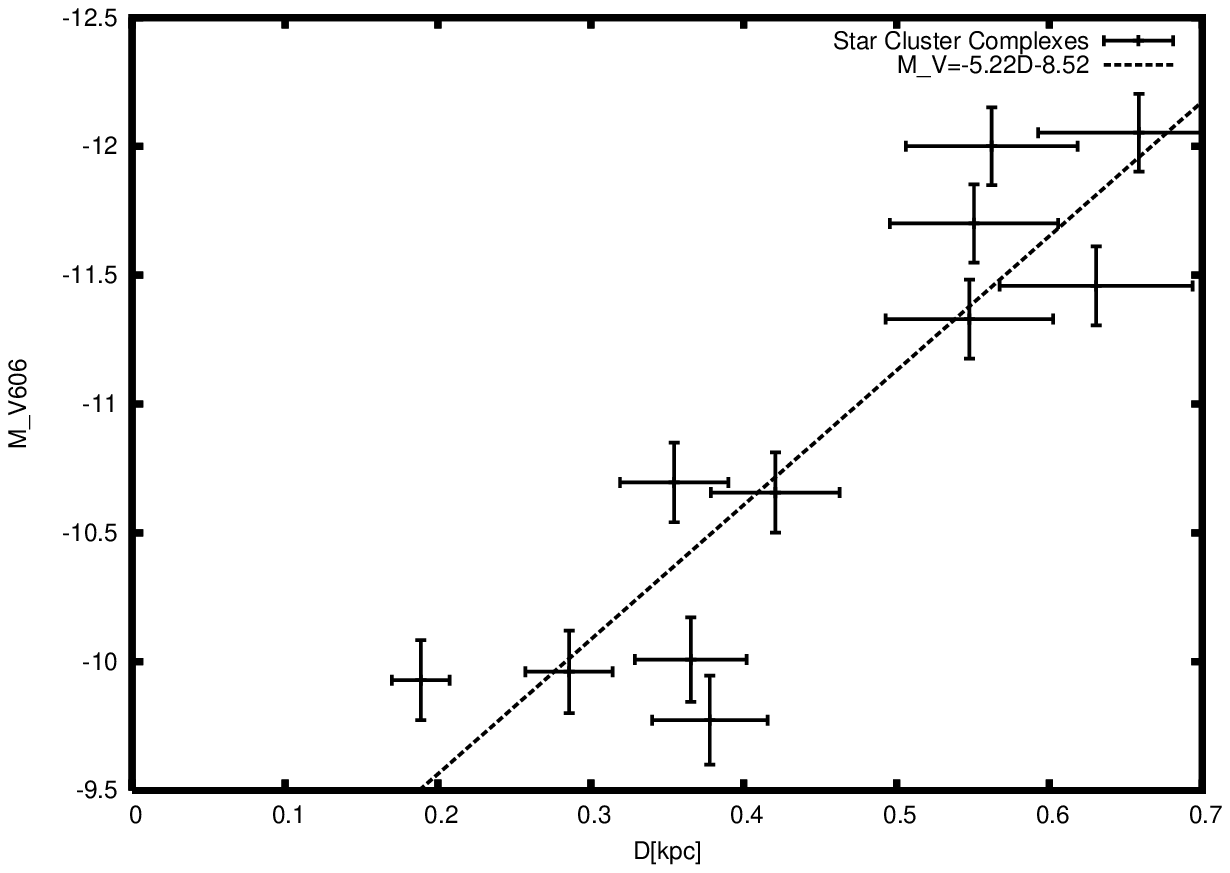}
\caption{ Left: HST/WFPC2 image in F606W of the Eastern tail of NGC 2782 from \citet{mullan} with star cluster complexes marked in red.  The area displayed here is the same as the inset in Fig.~\ref{fig:VimageE}.  Right: Size-luminosity diagram (D-M$_{V606}$) for star cluster complexes in the Eastern tail of NGC 2782.  We find the star cluster complexes to follow a linear size-luminosity relation similar to complexes found in nearby spirals and in HCG 59.  }
\label{fig:complex}
\end{figure}

Using the HST images from \citet{mullan} and the methods of \citet{isk7, isk59}, we find the boundaries of these extended amorphous structures by eye measured by contours which are $>10\sigma$ above the background (see Fig.~\ref{fig:complex}).  Photometry was performed on the polygonal apertures in the WF3 chip (the only one containing star cluster complexes in our tail region) using IRAF/POLYPHOT,  zeropoints from \citet{dolphin}, CTE correction for extended sources \citep{grogin}, and foreground extinction from \citet{schlafly} and NED for the HST filters.
We find 11 star cluster complexes in the Eastern Tail in the region between the main spiral and the putative dwarf galaxy.  
The size of each region was determined by taking the square root of the area contained in the polygonal aperture.  Since the boundaries of the complexes were determined by eye, we estimate the error to be $\sim10$\% of the size of the complex. The sizes of the star cluster complexes  ($0.19 < D < 0.63$ kpc) are consistent with sizes of star cluster complexes seen in nearby spiral galaxies \citep{elmsalzer} and in compact groups \citep{isk7,isk59}. As seen in Fig.~\ref{fig:complex}, there is a linear relation between size versus luminosity (M$_{V606}$) for star cluster complexes in the Eastern tail.  This indicates a uniform surface brightness for the star cluster complexes in the Eastern tail of NGC 2782.  The properties of the star cluster complexes are shown in Table~\ref{tab:complex} which lists: ID number, associated star cluster candidate  or {\HII} region ID (if any), centroid pixel values, photometry for F606W and F814W bands and associated errors, $V_{606}-I_{814}$ color, size of the region as calculated above, and $M_{V,606}$.  Of the 11 star cluster complexes, we find 6 of these have {\Ha} emission associated with them indicating ages $<10$ Myr.  All the complexes have blue color range of $-0.26< V_{606}-I_{814} <  0.65$ also consistent with previous observations of star cluster complexes with HST \citep{isk7,isk59}.

%Since tidal tail regions have different properties than spiral arms, this is unexpected.  

\begin{deluxetable}{cccccccccc}
\tabletypesize{\scriptsize}
\tablecaption{Properties of Star Cluster Complexes in the Eastern Tail of NGC 2782\label{tab:complex}}
%\vglue -0.5in
%\tablenum{4}
\tablewidth{6in}
\tablewidth{0pt}

\tablehead{\colhead{ID} & \colhead{SCCID} & \colhead{X} &\colhead{Y} &\colhead{Area} &\colhead{$V_{606}$} &\colhead{$I_{814}$} &\colhead{$V_{606}-I_{814}$} &\colhead{Size} &\colhead{$M_{V_{606}}$}  \\
\colhead{} &\colhead{} &\colhead{pixel} &\colhead{pixel} &\colhead{pixel$^2$} &\colhead{mag} &\colhead{mag} & \colhead{mag} &\colhead{kpc} & \colhead{mag}}

\startdata
1 & E0-A & 164.75 & 637.25 &   345.5 & 22.284(0.036) & 22.205(0.086) &  0.079(0.093) &  0.355(0.035) & -10.696(0.154)  \\
2 & E1-E & 123.90 & 259.50 &  1192.0 & 20.927(0.019) & 20.763(0.042) &  0.164(0.046) &  0.659(0.066) & -12.053(0.151)  \\
3 & E2-C & 231.50 & 432.33 &   833.5 & 21.279(0.022) & 21.057(0.047) &  0.222(0.052) &  0.551(0.055) & -11.701(0.152)  \\
4 & E3-F & 399.20 & 276.60 &   486.5 & 22.323(0.043) & 22.338(0.114) &  -0.015(0.122) &  0.421(0.042) & -10.657(0.156)  \\
5& E4-H & 240.83 & 200.50 &    98.0 & 23.052(0.041) & 23.310(0.128) &  -0.259(0.134) &  0.189(0.019) &  -9.928(0.156)  \\
6& E356 & 151.27 & 235.73 &  1093.0 & 21.522(0.031) & 21.083(0.054) &  0.439(0.062) &  0.631(0.063) & -11.458(0.153)  \\
7& E530 &  76.67 & 547.50 &   824.5 & 21.650(0.030) & 21.199(0.052) &  0.451(0.060) &  0.548(0.055) & -11.330(0.153) \\
8& E543 & 196.40 & 610.00 &   367.0 & 22.972(0.067) & 22.321(0.098) &  0.651(0.119) &  0.365(0.037) & -10.008(0.164)  \\
9& E555-B &  71.67 & 599.00 &   869.0 & 20.979(0.018) & 20.610(0.032) &  0.369(0.037) &  0.562(0.056) & -12.001(0.151) \\
 10 & & 181.40 & 136.40 &   224.5 & 23.019(0.056) & 22.571(0.097) &  0.449(0.112) &  0.286(0.029) &  -9.961(0.160)  \\
 11& & 226.75 & 655.50 &   392.5 & 23.207(0.085) & 23.169(0.216) &  0.038(0.232) &  0.378(0.038) &  -9.773(0.172)  \\

\enddata
\end{deluxetable}

\subsection{{\HII} Regions}
The {\Ha} observations yielded a detection of one source in the
Western tail (previously published in \citet{knierman12, werk, bournaud, tf12} and six sources in the Eastern tail (see
Figures~\ref{fig:VimageE} and \ref{fig:VimageW} for location of sources).
Previously, \citet{smith99} detected nine {\Ha} sources in the
Eastern tail.  Their observations were deeper than the ones presented
here.  Our detection limit for the Eastern tail is $7.7\times10^{36}$ erg \persec and for the Western tail is $3.7\times10^{37}$ erg \persec.  Table ~\ref{tab:clumps} lists the ID number of the star cluster candidate along with its letter designation from \citet{smith99} in the Eastern tail, tail inhabited, $B$ magnitude, $M_B$ magnitude, $U-B$, $B-V$, $V-R$, {\Ha} luminosity, and star formation rate (SFR) from the {\Ha} luminosity following \citet{kennicutt}.  
One of our most luminous {\Ha} sources, E4, is coincident with the location indicated for the tidal dwarf galaxy candidate in Fig. 2 of \citet{yoshida}.
In the Eastern tail, $L_{\Ha} = 1.8-7.5 \times 10^{38}$
erg s$^{-1}$, while the {\HII} region in the Western tail  (previously reported in \citet{knierman12}) 
has $L_{\Ha} = 19 \times 10^{38}$ 
erg s$^{-1}$.  The Western tail {\HII} region is listed in \citet{tf12} as System 6 which is the brightest of their young FUV sources in the Western tail.  
These luminosities are at least an order of magnitude less
than those of the brightest {\HII} regions in Sc galaxies ($L_{\Ha}
\sim 10^{40}$ erg s$^{-1}$), but are more typical of {\HII} regions in Sa/b galaxies
where the brightest ones are $L_{\Ha} \sim 10^{39}$ erg s$^{-1}$.  Both
tails have {\HII} regions fainter than 30 Dor which has $L_{\Ha} = 6
\times 10^{39}$ erg s$^{-1}$, but brighter than Orion ($L_{\Ha} =
10^{36}$ erg s$^{-1}$).  In comparison with the tidal arm region of NGC 3077 \citep{walter}, 
this very nearby tidal feature has 36 {\HII} regions with a total $L_{\Ha} = 2.9\times 10^{38}$ erg s$^{-1}$ which is similar to E3 and E4 and lower than W235.  
The {\Ha} region in the Western tail has blue colors and a young age fit with the 3DEF models.  Given their {\Ha} emission, these 7 sources are likely to be less than 10 Myr old indicating star formation in situ in both tidal tails.  

\begin{deluxetable}{ccccccccc}
\tabletypesize{\scriptsize}
\tablecaption{Properties of $H\alpha$ sources in Tidal Tails of NGC 2782\label{tab:clumps}}
%\vglue -0.5in
%\tablenum{4}
\tablewidth{6in}
\tablewidth{0pt}
\tablehead{\colhead{Number} & \colhead{Tail} & \colhead{B} & \colhead{$M_B$} & \colhead{U-B} & \colhead{B-V} & \colhead{V-R} &  \colhead{L$_{\Ha}$} & \colhead{SFR(\Ha)}\\
%\colhead{$M_{HI}$} & 
%\colhead{$M_{mol}$} \\
\colhead{} & \colhead{} & \colhead{mag} & \colhead{mag} & \colhead{mag} & \colhead{mag} & \colhead{mag} & \colhead{$10^{38}$erg s$^{-1}$} & \colhead{M$_{\odot}$ yr$^{-1}$} }
%\colhead{$M_{\odot}$} }
%\hline
\startdata
E0-A\tablenotemark{a} & E & 20.98 & -11.95 & -0.49 & 0.79 & 0.52  & $6.5(0.9)$  & 0.0051(0.0008) \\
E1-E\tablenotemark{a} & E & 21.57 & -11.37 & 0.12 & 0.23 & 0.38  & $7.5(1.0)$  &0.0059(0.0009) \\
E2-C\tablenotemark{a} & E & 21.17 & -11.76 & 0.01 & 0.17 & 0.59 & $1.8(0.4)$  & 0.0015(0.0003)  \\
E3-F\tablenotemark{a} & E & 21.60 & -11.33 & -0.20 & 0.51 & -0.28  & $2.7(0.5)$  & 0.0022(0.0004) \\
E4-H\tablenotemark{a} & E & 21.92 & -11.18 & -0.01 & -0.76 & 0.97  & $2.7(0.5)$  & 0.0022(0.0004) \\
E555-B\tablenotemark{a} & E & 20.996 & -11.994 & -0.138  &  0.417&  0.222 & $6.0(0.9)$  & $0.0047(0.0007)$ \\
%D\tablenotemark{a,b} & E & & & & & & & & & 1.2(0.3)& 0.0009(0.0002)\\ 
\\
\hline
\\
W235 & W & 22.492 & -10.498 & -0.60 & 0.42 & 0.33 & $19(3)$ & 0.015(0.002) \\
\enddata

(a) Letter Nomenclature from \citet{smith99}  
%\label{tab:clumps}
\end{deluxetable}

\subsection{Comparison to other star cluster populations in tidal debris}

In comparing NGC 2782 with the tidal debris of minor merger NGC 6872
\citep{bastian}, we find that both star cluster systems have similar
ages ($\sim10-100$ Myr).  However, NGC 6872 has more luminous star
clusters ($<M_{V}> = -12.2$) than NGC 2782 ($<M_{V,East}> = -11$;
$<M_{V,West}> = -10$). The difference in star formation can be seen in
comparisons of {\Ha} maps of both systems.  NGC 6872 has widespread
{\Ha} emission along both tails \citep{mihos93} which are correlated
with young massive star clusters \citep{bastian}.  On the other hand,
NGC 2782 has only a few discrete {\Ha} emission regions in
the Eastern tail, and only one in the Western tail.  NGC 6872 has a
slightly younger age ($\sim 145$ Myr) than NGC 2782 perhaps accounting
for the difference in star formation in the tidal tails.  There may
also be differences in the progenitor galaxies; these could account
for differences in star formation rate of the tidal tails.  NGC 2782
is a strongly star forming galaxy while NGC 6872 has little star
formation in its central regions.  The NGC 6872 interaction likely
stripped off gas-rich outer layers of the large spiral galaxy like the
Western tail of NGC 2782 while the Eastern tail of NGC 2782 may be
remnants of the smaller galaxy.
The sample of minor mergers by \citet{ferreiro}
include larger structures detected by {\Ha} emission, leading to
younger and more massive structures such as HII regions and Tidal
Dwarf Galaxies with very few star clusters detected.

The star clusters in the tidal tails of major merger NGC 3256
\citep{knierman} have similar ages (30-300 Myr) and luminosities
($<M_V> \sim -10$) to those in NGC 2782.  Three star clusters in the
Western Tail of NGC 3256 were spectroscopically confirmed by
\citet{trancho} at Gemini South.  From their GMOS-S spectra, they
determine ages to be $\sim80$~Myr for two star clusters and
$\sim200$~Myr for the third having approximately solar
metallicities. The three star clusters in the Western tail of NGC 3256
have masses ($1-2 \times 10^5$ \msun) which are similar to the average
masses of star clusters in the Western tail of NGC 2782.
Using the HST/WFPC2 images of the Western tail of NGC 3256,
\citet{trancho} show the three star clusters to have a large size
($r_{eff}\sim 10-20$~pc) compared to Milky Way globular clusters or
other young massive clusters ($r_{eff}\sim3-4$~pc).  The large size of
the star clusters may mean that the GMC underwent weak compression
during the star cluster formation.  On the other hand, the tidal tail
star clusters may not have experienced the tidal stripping that other
young clusters in the centers of galaxies have.

\subsection{Comparing Gas Properties of the Tails}
We now examine the gas properties of these two tails.  Both tails have been previously observed in {\HI} \citep{smith94} and in CO \citep{smith99, braine}.  The Eastern tail is rich in both {\HI} and CO while the Western tail has {\HI}, but no detectable CO.  Table~\ref{tab:sfrcomp} lists the location, area, SFR density ($\Sigma_{SFR}$) from {\CII}, SFR density ($\Sigma_{SFR}$) from {\Ha}, mass of {\HI}, molecular gas mass, total gas surface density, and predicted star formation rate density from the gas density from \citet{kennicutt}.    This table includes information for the local regions within each tail as well as for the entire tail.  

\begin{deluxetable}{ccccccccc}
%\begin{deluxetable}{lccccccccc}
\tabletypesize{\scriptsize}
\setlength{\oddsidemargin}{-0.5in}
\tablecaption{Comparison of Local \& Global Star Formation Rates}
\tablewidth{7.15in}
\tablewidth{0pt}
\tablehead{\colhead{Location} & \colhead{ID} & \colhead{Area} & \colhead{$\Sigma_{\rm SFR}(\CII)$} & \colhead{$\Sigma_{\rm SFR}(\rm H\alpha)$} & \colhead{M$_{\HI}$\tablenotemark{a}} & \colhead{M$_{\rm mol}$\tablenotemark{b}} & \colhead{$\Sigma_{\rm gas}$\tablenotemark{c}} & \colhead{$\Sigma_{\rm SFR}$(gas)\tablenotemark{d}}  \\

\colhead{} & \colhead{} & \colhead{${kpc^2}$} & \colhead{$10^{-3} M_\odot yr^{-1}kpc^{-2}$} & \colhead{$10^{-3} M_\odot yr^{-1} kpc^{-2}$} & \colhead{$10^8 M_\odot$} & \colhead{$10^8 M_\odot$} & \colhead{$M_\odot{\rm pc^{-2}}$} & \colhead{$10^{-3} M_\odot{ yr^{-1} kpc^{-2}}$}}
\startdata
\bf{East} & & & & & & & &  \\
TDGC & E1,E4 & 7.09 & $4.5(0.2)$ & $1.1(0.1)$ & $0.4$ & $2.6$\tablenotemark{f} & $44$ & 51 \\
\HI-N & E3 & 11.8 & $<0.3$ & $0.27(0.04)$ & $5.2$ & $2.6$\tablenotemark{f} & $82$ & 119 \\
\HI-M & E2 & 7.09 & $2.5(0.3)$ & $0.22(0.05)$ & $3.1$ & $3.69(0.09)$ & $112$ & 184\\
\HI-S & E0,E555 & 23.6 & $2.9(0.1)$ & $0.42(0.03)$ & $8.1$ & $3.8$\tablenotemark{f} & $62$ & 82 \\
\hline 
E Tail &  & 3000 & 0.04 & 0.01& 17& 8.6& 1.1& 0.3 \\
\hline
\hline\\
\hline
\hline
\bf{West} & & & & & & &   \\
\HI-N & & 8.6 & & $<0.03$ & $0.73$ & $<0.086$\tablenotemark{e} & $<12.5$ & $<9$ \\
\HI-M & W235& 14.7 & $<1.1$ & 1.0(0.2) & $1.15$ & $<0.22$& $<9.1$ & $<5$ \\
\HI-S & & 19.3 &  & $<0.02$ & $1.16$ & $<1.5$\tablenotemark{f} & $<10.1$ & $<6$ \\
\hline 
W Tail &  & 2300 & & 0.009 & 19 & $<1.8$ & $<1.6$ & $<0.5$ \\

\enddata
\tablenotetext{a}{\citet{smith94}, corrected for distance}
\tablenotetext{b}{M$_{\rm mol}$ inferred from CO observations}
\tablenotetext{c}{Includes helium (M$_{\rm gas} = 1.36(\rm M_{\HI} + M_{\rm H_2}$))}
\tablenotetext{d}{From \cite{kennicutt},  $\Sigma_{\rm gas}$ includes only {\HI} and H$_2$}
\tablenotetext{e}{\citet{braine}, corrected for distance}  
\tablenotetext{f}{\citet{smith99}, corrected for distance, TDGC and E-\HI-N use Far-Far-East pointing while E-\HI-S uses South-Far-Far-East pointing.}

\label{tab:sfrcomp}

\end{deluxetable}

\subsubsection{Neutral Hydrogen - {\HI}}

There are 3 massive {\HI} clumps in both tails which have star cluster
candidates associated with them.  In the Western tail,
\citet{smith94} measure 10 massive {\HI} clumps.
These range in masses from $3\times10^7$ \msun to $1.8\times10^8$
\msun.  Only three of these massive {\HI} clumps have star cluster
candidates found within their bounds.  Since only one {\HII} region was
found in the Western tail, we use the {\Ha} background limit as a limit
for the {\Ha} star formation rate for the other two {\HI} clumps.  The
magenta boxes in Fig. \ref{fig:VimageW} show the location of
these three clumps in the Western tail.  In the Eastern tail,
individual clumps are not tabulated in \citet{smith94}, but their
masses and sizes were determined by inspection of their map of {\HI}
contours.  We also include a box of the area surrounding the tidal dwarf galaxy candidate (TDGC) found by \citet{yoshida} even though it falls outside the {\HI} peaks.  The {\HI} clumps in the Eastern tail have masses from $3-8\times10^8$ \msun, 
but only $4\times10^7$ {\msun} for the location of TDGC.
The four regions in the Eastern tail have {\Ha} sources
which provide the local {\Ha} luminosities.  If more than one {\HII} region resides in a box, 
we sum the luminosities of the regions.  The total {\Ha} luminosity for each region is presented in 
Table~\ref{herschel}.  We find that TDGC now has the highest $L_{\Ha}$ in the Eastern tail, however, the
Western tail {\HII} region remains the brightest.  
The magenta boxes in
Fig. \ref{fig:VimageE} show the location of these four areas in
the Eastern tail.  

\subsubsection{Cold Molecular Gas - CO}

\citet{smith99} used the Kitt Peak 12 meter to look for CO(1-0) in NGC 2782.  They used a grid pattern with 25\arcsec \ spacing and detect CO in 5 out of 6 pointings in the Eastern tail region.  These locations are shown as asterisks in Figure~\ref{fig:contour}.  Only 2 pointings (Far-Far-East and South-Far-Far-East) are located in the same region as the Herschel PACS observations.  To correlate the Herschel PACS observations more accurately with molecular observations, we made new observations with the Kitt Peak 12 meter at the same location.  The Eastern tail location was detected at a lower signal-to-noise (about 4$\sigma$, see Fig.~\ref{fig:CO}) than previous observations, but has the same velocity as the areas observed in the Eastern tail by \citet{smith99}.  Assuming the source fills the beam (a coupling efficiency of $\eta_c = 0.64$) and a Milky Way $X_{CO}$, the molecular mass observed at the same location as the Herschel PACS observations is M$_{mol} = 3.69\pm0.09\times 10^8$ \msun.  

In the Western tail, \citet{smith99} looked for CO at two locations, the northern and southern {\HI} clumps, but did not detect any CO(1-0) emission.  The {\HII} region in the Western tail \citet{knierman12} is located $\sim20$\arcsec  \ away from the location where \citet{smith99} searched for CO(1-0), so we made deeper observations at this location. The Western tail location, coincident with the {\Ha} source \citep{knierman12}, remains undetected in CO(1-0) with an upper limit of M$_{mol} < 0.22\times 10^8$ \msun.  
\citet{braine} reobserved the northern clump with IRAM, but still did not detect CO.  Table~\ref{tab:sfrcomp} lists the molecular gas mass or upper limits from these pointings.  

The use of the standard Milky Way $X_{CO}$ is not without precedence for tidal debris regions even though major merger TDGs have been observed to have $\sim0.3Z_{\odot}$ since they are pulled from the outer regions of spiral galaxies \citep{duc}.  The previous observations of the tidal debris of NGC 2782 by \citet{smith99} use the Milky Way conversion factor for ease of comparison with other observations.  In \citet{boquien} and \citet{walter}, the standard Milky Way CO to  H$_2$ conversion factor (X$_{CO} = 2 \times 10^{20}$ cm$^{-2}$ (K km s$^{-1}$)$^{-1}$ or ($\alpha_{\rm CO 1-0} = 4.3 $  \msun (K km s$^{-1}$ pc$^2)^{-1}$ ) was used.  However, as mentioned in \citet{knierman12}, $X_{CO}$ has  a strong dependence at low metallicities \citep{leroy, genzel}.  At a metallicity of 0.3$Z_{\odot}$ (or $\mu_0 = 8.19$), the  conversion factor is $\alpha_{\rm CO 1-0} = 27.5$  \msun (K km s$^{-1}$ pc$^2)^{-1}$ and gives a factor of 6 higher molecular mass limit  
 (M$_{\rm mol} \leq 1.2 \times 10^8$ \msun) than that from the 
standard conversion factor (M$_{\rm mol} \leq 2 \times 10^7$ \msun).

\subsubsection{Herschel/PACS observations of {\CII}}
Observations of {\CII} with Herschel/PACS show two major peaks in the Eastern tail (Fig.~\ref{fig:contour}), but no
detection in the Western tail.  The peaks in the Eastern tail correspond with the two strongest {\Ha} sources, E0 and E1.  The northern peak in {\CII}  (TDGC) is near E1 (the location indicated for the TDGC \citep{yoshida}) and E4.  The southern peak (E-HI-S) is near the brightest {\Ha} source, E0, and the {\HI} peak in the Eastern tail.  A secondary peak of {\CII} in this area is near the {\Ha} source E555-B.  The locations where spectra were extracted are indicated in Figure~\ref{fig:contour} with the spectra shown in Figure~\ref{fig:spectra}.   As above, the extracted spectra were summed, multiplied by the bandwidth (745.96 MHz) and the flux correction factor for a single pixel (2.0), and converted to cgs units by the conversion factor of $10^{-23}$ erg \persec cm$^{-2}$.  We multiply the flux by $4\pi D^2$ to calculate the luminosity.  To calculate I$_{\CII}$, we divide the flux by the circular beam size of 12\arcmin.  For each location, Table~\ref{herschel} lists the location, coordinates, luminosity of {\CII}, luminosity of {\Ha}, I$_{CO}$ from this work or for the nearest location observed by \citet{smith99},    I$_{\CII}$/I$_{CO}$, the SFR from {\CII} calculated using the relation from \citet{boselli}, and SFR from {\Ha} using \citet{kennicutt}.  Included in the table are limits for the regions with no strong detections.  

The southern source, E-HI-S, is 1.1 times brighter than the TDGC source and has a luminosity of $15.1\pm0.6\times10^{38}$erg s$^{-1}$.  E-HI-S is similar in brightness to its neighbor, B, though some of the {\CII} flux near B could be off the array.  So this observation may be a lower limit.  We extract a spectrum at the location of the {\Ha} source, E2-C or E-HI-mid, even though no peak is observed in the line intensity map.  Since there is {\CII} emission there at a level of 4-5$\sigma$, we detect a line with luminosity less than half the value of E-HI-S.   We also extract a spectrum near the location of the northeast {\Ha} source, E3/E-HI-N, where there is no {\CII} emission detected.  
The Western tail has an upper limit of $<0.6\times 10^{38}$erg s$^{-1}$.  

To calculate the SFR from {\CII}, we need to correct the line luminosity to account for contamination from the warm ionized medium.  \citet{boselli} cite that $2/3$ of the {\CII} flux comes from the neutral medium based on theoretical predictions.   Based on models of {\HII} regions, \citet{mookerjea} use 70\% of the {\CII} from the neutral medium for their calculations which is consistent with their non-detection of the [NII] 205$\mu$m line from BCLMP 302 in M33.  The first observational determination of the fraction of {\CII} flux from the neutral medium  is from \citet{oberst}.  They use the first detection of  the [NII] 205$\mu$m line and show that 27\% of the {\CII} line flux should come from the warm ionized medium, leaving 73\% of {\CII} to come from the neutral medium.  While the tidal tails of NGC 2782 may be different from the Carina nebula observed by \citet{oberst},  we adopt the \citet{oberst} value for our determination of SFR from {\CII} emission due to the paucity of observations of the [NII] 205$\mu$m line.  
To calculate SFR from {\CII}, we multiply the {\CII} flux by 0.73 \citep{oberst} and use Eq. 3 from \citet{boselli} who observe nearby late-type galaxies in {\Ha} and {\CII}.  

\begin{deluxetable}{cccccccccc}
\tabletypesize{\tiny}
\setlength{\oddsidemargin}{-0.5in}
\tablecaption{{\CII} and {\Ha} Observations of Regions in Eastern and Western Tails}
\tablewidth{7.15in}
\tablewidth{0pt}
\tablehead{\colhead{Location} & \colhead{\Ha ID} & \colhead{RA} & \colhead{Dec}  & \colhead{L$_{\CII}$ } & \colhead{L$_{\Ha}$} & \colhead{I$_{CO}$} & \colhead{I$_{\CII}/$I$_{\rm CO}$\tablenotemark{b}} & \colhead{SFR(\CII)\tablenotemark{b}} &\colhead{SFR(\Ha)\tablenotemark{c}}\\
\colhead{} & \colhead{} & \colhead{} & \colhead{} & \colhead{$10^{38}$ erg s$^{-1}$} & \colhead{$10^{38}$ erg s$^{-1}$}  & K km s$^{-1}$ & &  \colhead{\msun yr$^{-1}$}& \colhead{\msun yr$^{-1}$}}
\startdata
\bf{East} & & & & &\\
TDGC & E1,E4 &9:14:10.92 & +40:06:48.8 & 13.4(0.6) & 10(1.2) & 0.39(0.06)\tablenotemark{a} & 2950 & 0.032(0.001) & 0.008(0.001)\\
E-HI-N  & E3 & 9:14:12.46 & +40:06:50.6 & $<$0.9 & 4.1(0.6) & $0.39(0.06)\tablenotemark{a}$ & $<$190& $<$0.004 & 0.0032(0.0005)\\
E-HI-M & E2 & 9:14:11.46 & +40:06:35.6 & 6.3(0.7) & 2.0(0.4) & 0.43(0.05) & 1840 & 0.0170(0.005) & 0.0016(0.0003)\\
E-HI-S & E0 & 9:14:11.26 & +40:06:20.6 & 15.1(0.6) & 6.7(0.9)& 0.57(0.08)\tablenotemark{a} & 4860 & 0.0347(0.001) & 0.0053(0.0007)\\
E-HI-S-B & E555 & 9:14:10.66 & +40:06:26.6 & $>$14.1(0.7)\tablenotemark{f} & 6.0(0.9) & 0.57(0.08)\tablenotemark{a} &4520 & $>$0.033(0.002)& 0.0047(0.0007)\\
\hline
%FFE\tablenotemark{e}  & E1,2,3,4& 9:14:11.63& +40:06:48.6& 6.1(0.2)& 15(2)& 0.39(0.06) & 93 & 0.0169(0.0005) & 0.012(0.001) \\
%SFFE\tablenotemark{e}& E0,2,B & 9:14:11.63& +40:06:23.6& 6.8(0.2)&  15(2) & 0.57(0.08) & 71& 0.0184(0.0005) & 0.012(0.002) \\
Total\tablenotemark{e} & all & 9:14:10.63 & +40:06:41.6 & 77(2) & 29(2) & 0.43(0.05) & $>1148$ & 0.125(0.003) & 0.023(0.002)\\
\hline
\hline
\bf{West} & & & & & \\
W-HI-M & W235&9:13:51.2 &+40:08:07 & $<$5.6& 19(3) & $<$0.02 & $<$35200&$<$0.02 & 0.015(0.002)\\

\enddata
%\begin{flushleft}
\tablenotetext{a}{\citet{smith99}}
\tablenotetext{b}{1 K km s$^{-1} = 1.6 \times 10^{-9}$ erg s$^{-1}$ cm$^{-2}$ sr$^{-1}$ \citep{stacey91}}
\tablenotetext{c}{Using equation from \citet{boselli}}
\tablenotetext{d}{Using equation from \citet{kennicutt}}
\tablenotetext{e}{Total of the area of the Herschel/PACS spectral observations.}  
\tablenotetext{f}{This is a lower limit since source is at the edge of the array.}
%\end{flushleft}

\label{herschel}

\end{deluxetable}

\section{Discussion}
Based on the observations described above, we find both tidal tails of NGC 2782 host
young star forming regions that formed within the tidal tail.  The Eastern tail has more luminous star clusters which are hosted in larger star cluster complexes whereas the Western tail has only isolated star clusters.  Therefore, packaging of star formation is different between the tails.  The Eastern tail also has CO and {\CII} emission, whereas the Western tail has non-detections.   The {\HII} region in the Western tail is more luminous than any single {\HII} region in the Eastern tail, so the difference between the tails is not simply that the Western tail is forming stars at a lower level than the Eastern tail.
To determine what might be causing these differences between two tidal tails of the same system, we compare the following properties: ambient pressure, gas phase, amount of gas, and efficiency of star formation.

\subsection{Ambient Pressure}

The difference in star formation modes between the tails could be due 
to a differing initial distribution of star cluster masses.  Differences in 
cluster initial mass function (CIMF) are difficult to determine since 
it is hard to find young star cluster populations that have not experienced
significant evolution.  A few studies have attempted determining the CIMF
\citep{review}.  The CIMF is generally described like a Schechter distribution

\begin{equation}
\frac{dN}{dM} = AM^{-\beta} exp(-M/M_*)
\end{equation}

with $\beta \sim 2$ and $M_*$ indicating the mass at which the  change in the slope
of the mass function occurs.  For spiral galaxies like the Milky Way, $M_* \sim 2 \times
10^5$ \msun, but interacting galaxies like the Antennae show $M_* > 10^6$ \msun \citep{review}.  
Therefore, the environment where star cluster formation occurs seems to affect the CIMF and, in particular, the value of $M_*$.  

This possible difference in CIMF between the tails might be due
to the difference in the spatial arrangement of molecular gas. Forming
stars tend to destroy their parent molecular environment, leaving
behind clumpy remnants of molecular material. Beam dilution may render
the CO unobservable even with moderate resolution ($<1^{\prime}$) like 
the CO observations of the {\HI} peaks in the Western tail \citet{smith99} and \citet{braine}.
With the concentration of molecular gas in small clumps, the tail
may have produced correspondingly lower mass star clusters.  

A second possibility for the difference in CIMF could be due to 
a lower metallicity.  With a lower metallicity there would be less carbon
and oxygen to make CO so we may not detect it.  However, why a lower
metallicity environment would make lower mass clusters is an open
question. 

A third possibility for the difference in CIMF between the tails could be due to the environment.  The ambient pressure in the Western tail could be lower than in the Eastern tail
which would lower the star formation rate (Blitz \& Rosolowsky 2006),
possibly making lower mass star clusters.  Even for solar
metallicities, when the ambient pressure is low, CO will form deeper
within the dense star forming cloud.

We examine the third possibility next.  
The locations and masses of GMCs are likely to be regulated by the
structure of the {\HI} from which the GMC formed, particularly in
galaxies which are dominated by atomic gas \citep{blitz06}.  In
particular, M33 shows this correlation between its GMCs and peaks in
the {\HI} gas \citep{rosolowsky07}.  

While \citet{braine} stated that the lack
of CO down to sensitive limits at places of high column density {\HI} indicated that the Western tail is not
gravitationally bound and so the gas has not collapsed to form H$_2$
and hence new stars, the observations presented here show that new
star clusters are being formed in the tail.
Based upon {\HI} peaks in the Western tail of NGC 2782, GMCs are expected
to be at those locations.  The largest size of the molecular cloud that one would expect to be
associated with the {\HI} clump can be estimated using the formulation
for $M_{char}$ of \citet{elmegreen93}:

\begin{equation}
M_{char} = \pi l_{min} c_g \sqrt{\frac{\mu}{2G}}
\end{equation}

where $c_g$ is the velocity dispersion of the gas in 3-D, $\mu$ is the
mass per unit length of the cloud ($M/l_{max}$), and $l_{min}$ and
$l_{max}$ are the minor and major axes of the {\HI} cloud.  This equation
gives the characteristic mass of a ``supercloud'' using the size,
mass, and velocity dispersion of the {\HI} clump.  This supercloud is the
overall entity which becomes self-gravitating and the GMCs collapse to
form inside this supercloud by turbulent fragmentation. 

Table~\ref{tab:mchar}  has the $M_{\HI}$, velocity dispersion, size and resulting
$M_{char}$ for 3 {\HI} clumps in the Eastern tail and 3 {\HI} clumps in the
Western tail.  \citet{smith94} measured 10 {\HI} clumps in the Western
tail, but the 3 presented here are ones with star cluster candidates
at that location.  The $M_{char}$ for the Eastern Tail clumps ranges
from $9.5-18 \times 10^8$ \msun while the Western tail is slightly
smaller with a range from $6.6-11 \times 10^8$ \msun.  

%\clearpage
\begin{deluxetable}{ccccccc}
\tabletypesize{\scriptsize}
\setlength{\oddsidemargin}{-0.5in}
%\begin{table}[ht]
\tablecaption{Characteristic Sizes of HI Clumps in the Tidal Tails of NGC 2782}
\tablewidth{7.15in}
\tablewidth{0pt}
%\begin{tabular}{ccccccc}
%\tabletypesize{\footnotesize}
%\vglue -0.5in
%\tablenum{4}
%\tablewidth{6in}
%\tablewidth{0pt}
\tablehead{\colhead{Location} & \colhead{$\sigma_{v}$\tablenotemark{a}} & \colhead{$l_{min}$} & \colhead{$l_{max}$} & \colhead{$M_{HI}$\tablenotemark{a}} & \colhead{$M_{char}$} & \colhead{Observed $M_{mol}$}\\
%\hline
\colhead{}    & \colhead{(km/s)} & \colhead{(kpc)} & \colhead{(kpc)} & \colhead{($10^8 M_{\odot}$)} & \colhead{($10^8 M_{\odot}$)} & \colhead{($10^8 M_{\odot})$}}
\startdata
E-HI-N & 30 & 2.96 & 2.96 & $3.85 $ & $15.3$ & $2.6(0.5)$\tablenotemark{b}\\
E-HI-M & 40 & 1.78 & 2.96 & $2.31$ & $9.50$ & $3.69(0.09)$\\
E-HI-S & 40 & 2.96 & 5.92 & $6.02$ & $18.1$ & $3.8(0.5)$\tablenotemark{b}\\
\hline
% & & & & & & \\
\hline
W-HI-N & 49 &  2.10 & 3.05 & $0.54 $ & $6.56 $ & $<0.086$\tablenotemark{c}\\
W-HI-M & 35 &  5.67 & 7.56 & $1.00$ & $10.9$ & $<0.22$\\
W-HI-S & 30 &  3.78 & 3.78 & $0.86$ & $8.19$ & $<1.5$\tablenotemark{b}\\
\hline
 %& & & & & & \\
\hline
%HI clump & $\sigma_{v}$ & $l_{min}$ & $l_{max}$ & $M_{HI}$ & $M_{char}$ & Observed $M_{mol}$\\
%\hline
\hline
N7252 TDG & 31 &  5.00 & 5.00 & $10$ & $33.2$ & 0.2\\
%\hline
\enddata

\tablenotetext{a} {\citep{smith94, smith99}}
\tablenotetext{b} {\citep{smith99}}
\tablenotetext{b} {\citep{braine}}

\label{tab:mchar}
%\end{tabular}
%\end{table}
\end{deluxetable}

While $M_{char}$ does not predict an exact mass for a GMC, this does
indicate that similar masses of GMCs are expected in both the Eastern
and Western tails.  However, no molecular gas was observed in the
Western tail down to a limit two orders of magnitude below $M_{char}$.
A massive {\HI} clump in the NW tidal tail of NGC 7252 has $M_{char} \sim
3.3 \times 10^9$ \msun.  This clump, as shown in Table~\ref{tab:mchar}, has more HI
than those in NGC 2782 and molecular gas of $2 \times 10^7$ \msun.  

That the Western tail lacks molecular gas at the locations of high {\HI} column density
could be due to either its
absence or its inobservability.  In the less dense region of the
Western tail, even a small amount of star formation could have
destroyed the molecular gas there.  The Eastern tail is more dense and
so its molecular gas would not be destroyed as easily.  If the
molecular gas is not entirely destroyed by the radiation of young
stars, it could exist instead in smaller clouds which would be greatly
affected by beam dilution with single dish telescopes (beam sizes of
 22\arcsec \ to 1\arcmin).

\subsection{Gas Phase}
We examine whether the lack of massive star clusters and star cluster complexes in the Western tail is related to the properties of the ISM in the 
tidal tail.  To do this, we look at various tracers of the reservoir of gas available for star formation in both tails on both local and global scales.  We first compare neutral hydrogen gas to molecular gas traced by CO emission.  Then we compare neutral hydrogen to ionized gas traced by {\CII} emission.  Finally, we compare molecular gas traced by CO emission to ionized gas traced by {\CII} emission.

Comparing the CO and {\HI} mass ratio, we find the highest to be TDGC ($M_{H_2}/M_{\HI} = 6$) the other three regions have ratios of 0.5, 1, and 0.5 for the north, mid, and south regions, respectively.  In the Western tail, we find 
$M_{H_2}/M_{\HI} < 0.2$.  This indicates that there is less CO in the Western tail for the {\HI} mass as compared to the Eastern tail on local scales.

We next compare the {\CII} to {\HI}.  In Fig. 16 of \citet{stacey91}, they plot the expected {\CII} line flux for given {\HI} column densities for various phases of the ISM in the infinite temperature limit.  The galaxies of their sample lie about 2 dex above the lines for ``standard" {\HI} clouds and the intercloud medium.  In direct comparison to their galaxies and Galactic {\HII} regions, the local regions in the Eastern and Western tail of NGC 2782 are deficient in {\CII} by about 2 dex.  The Eastern tail has an average {\HI} column density of $6\times10^{20}$ cm$^{-2}$, with a peak of  $1.75\times10^{21}$ cm$^{-2}$ in the E-HI-S region.  For a ``standard" {\HI} cloud ($n_H \sim 30$ cm$^{-3}$), the peak column density in the Eastern tail predicts $I_{\CII} \sim 5\times10^{-6}$ erg s$^{-1}$ cm$^{-2}$ sr$^{-1}$ which is a factor of 1.6 larger than that observed in E-HI-S.  If we consider the average {\HI} column density for the Eastern tail, then $I_{\CII} \sim 2\times10^{-6}$ erg s$^{-1}$ cm$^{-2}$ sr$^{-1}$ which is similar to the values observed for the Eastern tail.  In the Western tail, the {\HI} column density is $\sim1\times10^{21}$ cm$^{-2}$ and the predicted value for $I_{\CII}$ is a few times the observed upper limit for {\CII}.  

To directly compare {\CII} to CO, we determine the ratio of their intensities.  Our values for $I_{\CII}$ are calculated by dividing the flux of {\CII} by the beam size (a HPBW of 12\arcsec).  
As shown in Table~\ref{herschel}, we observe the $I_{\CII}/I_{CO}$ ratio to range from 1840-4860 for those where both {\CII} and CO are detected.  This is at the lower end of the values for most of the normal and starburst galaxies observed \citet{stacey91} and aligns more with the values for Galactic {\HII} regions and molecular clouds.  \citet{stacey91} found that there was a constant ratio between {\CII} and CO fluxes for starburst and Galactic OB star forming regions ( $I_{\CII}/I_{CO}= 6300$).  The ratio for TDGC is 2950 and the ratio for E-HI-S is 4860.  These values are a few times lower than the constant ratio.  The limit for the Western tail gives a  $I_{\CII}/I_{CO} > 35200$.  \citet{stacey91} finds the ratio to be lower for non-OB star forming regions in the Milky Way and normal galaxies with lower dust temperatures and those ratios range from 4400 to as low as 900 which match with our observations in the Eastern tail.  Our limit in the Western tail is larger than the constant ratio from \citet{stacey91} and is more similar to their value for 30 Dor in the LMC ($I_{\CII}/I_{CO} = 40,000$) which is also an area of lower metallicity than most of their other observed regions.  If we compare to the more recent observations of area of the {\HII} region BCLMP 302 in M33 \citet{mookerjea}, we find our values align to their range of ratios from  $1000-70,000$ shown in Fig. 12 of their erratum.  For the {\HII} region BCLMP 302 itself, the values of $I_{\CII}/I_{CO}$ range from $6000-30,000$ which are higher than our ratios for the Eastern tail, but have an upper end similar to the limit for the Western tail {\HII} region.  The {\HII} region BCLMP 302 in M33 has a $L_{\Ha} = 2.2\times 10^{38}$ erg \persec which is similar to the {\HII} regions in the Eastern tail.  However, the {\CII} intensity in BCLMP 302 is 28 times lower in the Eastern tail and 84 times lower than the Western tail limit.  

There may be a mismatch between the CO observations and the {\CII} peaks from the different beam sizes of the two instruments.  PACS spectrometer has a beam with half power beam width (HPBW) of 12\arcsec \ at 159$\mu$m while the CO data are from the Kitt Peak 12 meter which has a beam with FWHM of 55\arcsec.  We instead extract {\CII} spectra from the PACS array by summing the flux from the whole array to better match the CO beam and by dividing the flux by the area of the array ($47\arcsec \times 47\arcsec$).  This is indicated in Table \ref{herschel} as ``Total".  Even with summing the flux from the whole PACS array, the CO observations still have a larger beam size which means we have a lower limit on the ratio of $I_{\CII}/I_{CO} > 1148$.  

In either case, it appears that the Eastern tail has a deficiency of {\CII} since the $I_{\CII}/I_{CO}$ ratio is on the low end of regions previously observed.  And given that this is among the first tidal tail region to be observed in {\CII}, having a lower value for $I_{\CII}/I_{CO}$ may not be surprising.

\subsubsection{Non-detection of {\CII} in the Western tail}
Given the non-detection of CO in the Western tail, we might expect the molecular gas to be in another form.
CO does not form until and $A_V$ of 3 or more, while H$_2$ forms at $A_V$ of less than 1 \citep{ht97}. In a
low pressure environment like a tidal tail a substantial amount of molecular gas can
exist with conditions that do not favor the formation of CO, so the CO
to H$_2$ conversion factor is not a constant. In the low gas density
environment of tidal debris a substantial reservoir of molecular gas
can exist at low $A_V$ that will not be detectable through CO.  However, 
in this regime, C$^+$ will be present in higher amounts.   Theoretical models for molecular clouds in \citet{wolfire} show that the fraction of molecular mass in the  ``dark gas" (H$_2$ and C$^+$) is $f \sim 0.3$ for typical galactic molecular clouds.   For lower $A_V$ and lower metallicities (as in these tidal tail regions), the fraction of dark mass in H$_2$ increases.  Given this reason, we observed the Western tail with Herschel to see if {\CII} was present where CO was not.  

We do not find {\CII} in the Western tail at the location of the {\HII} region at a significant level.  This {\HII} region
has a higher {\Ha} luminosity than any individual {\HII} region in the Eastern tail, but {\CII} emission 
is only detected in the Eastern tail.   Given that {\Ha} and {\CII} emission ultimately originate from the same sources of hard UV photons with energy greater than 13.8 eV, 
they are expected to trace each other.   In BCLMP 302 in M33, \citet{mookerjea} find {\CII} emission 84 times greater than the Western tail limit, even though it has an {\Ha} luminosity 9 times less than the Western tail.

If carbon was not available to be ionized, the most likely place for it to be is in CO.  However, the Western tail has no detected CO at this location.  From the calculations of the characteristic mass in Section 4.2 of the paper, we expect the largest size of the molecular cloud mass to be $\sim1\times10^9$ \msun, but the upper limit for this region is $0.2\times10^8$ \msun.  (in the Eastern tail, M$_{char}$ is $\sim3-5$ times the observed value of M$_{mol}$, so the factor of 54 between M$_{char}$ and the upper limit on molecular mass in the Western tail is significant). 
The lack of {\CII} and CO emission would then indicate that this {\HII} region is likely to have a low carbon abundance.  However, previous emission line spectra of this {\HII} region \citep{tf12, werk} from Gemini have shown that it has a metallicity greater than solar.  This seems to be at odds with the non-decections of {\CII} and CO in the Western tail.  However, \citet{tf12} use nebular oxygen emission for their metallicity determination.  
It is possible for this to be compatible with a low carbon abundance and a very low C/O ratio.  Oxygen is the most common element produced in core collapse supernovae events whereas carbon is produced in relatively small amounts.  The source of most carbon in the ISM is AGB stars.  If this material has been primarily enriched by recent star formation, it is possible to build up a high abundance of oxygen and alpha elements without producing a significant enhancement of carbon and also iron.  The age of the Western tail is about 200-300 Myr and so in situ star formation in the tail is unlikely to have synthesized large quantities of carbon \citep{arnett96}.  

Also, molecular gas may have been dissociated by a high UV flux in this region, associated with the most massive stars in the star cluster.  FUV-NUV color from \emph{GALEX} observations by \citet{tf12} show that this region show a FUV-NUV color of -0.14 mag. 
For a simple Salpeter IMF of $dN/dM = cM^{-(1+x)}$ with
x=1.35, a $10^{5.35}$ \msun cluster contains 1660 stars of 8 $M_\odot$ or
greater. Even for an extreme value of x=2.35,
there are still 40 stars of 8 or more $M_\odot$. 
This provides
sufficient UV flux to photodissociate H$_2$ out to several hundred
parsecs, comparable to the size of the Orion Molecular Cloud
Complex. It is therefore unlikely that molecular gas has long
lifetimes unless present at high densities, as in the Eastern tail of
NGC 2782.

\subsection{Star Formation in Tidal Debris}
To consider whether the difference in star formation modes between the
two tails is due to the amount of gas present for star formation, in
each tail we calculate the star formation rates per unit area from the
{\Ha} luminosity and also from {\CII} luminosities and compare these to the predicted SFR from the gas surface
density.  This is done for global values of the tidal tails as
well as locally over {\HI} clumps in both tails.

\subsubsection{Star Formation on Global Scales}
Using the entire tail areas as calculated above, we find the SFR and SFR per area ($\Sigma_{SFR}$) for both tails.  
In Table~\ref{tab:sfrcomp}, the global properties of
$\Sigma_{SFR}$ from {\CII}, from the {\Ha} luminosity, and the total gas surface
density (including both {\HI} and molecular gas) are listed for each
tail.  For this calculation we use the L$_{\Ha}$ from
\citet{smith99} since they detected fainter {\Ha} emission than these
observations.  In the Eastern tail the $\Sigma_{SFR} (\Ha)$ is an
order of magnitude below the expected value of $\Sigma_{SFR}$(gas).  
In the Western tail, $\Sigma_{SFR} (\Ha)$ 
is two orders of magnitude below the expected value from the gas density.

Using the summed flux over the entire PACS spectroscopy array, as described above and equation for SFR from {\CII} in \citet{boselli}, 
we calculate the Global SFR (\CII) for the Eastern tail to be 0.125 {\msun} yr$^{-1}$.  If we 
assume that the flux detected by the PACS pointing is the entirety of the {\CII} in the Eastern tail, 
$\Sigma_{SFR} = 4 \times 10^{-5}$ \msun yr$^{-1}$ kpc$^{-2}$.  This value is 4 times larger than 
the value from $\Sigma_{SFR} (\Ha)$, but still 8 times less than the expected value from the 
gas density.  However, \citet{boselli} find that the dispersion in their SFR correlation $\sim10$.  So a
factor of 4 difference is within the scatter.  They do say that {\CII} line luminosity can be taken as a star formation indicator for normal late-type galaxies ($8.0 < \log L_{FIR} < 10.5$), but may be not applicable to ULIRGs.  The NGC 2782 galaxy itself  is a LIRG with an IRAS luminosity of $1.6\times10^{44}$ erg s$^{-1}$ which is at the high end of this range.  However, we study the tidal tails which would have lower FIR emission than the central regions of the merger.  Whether to use {\CII} as a direct indicator for star formation is a matter of some debate, however, there have been recent calibrations of the SFR from {\CII} for different regimes such as starbursts \citep{delooze} and {\HII} regions in M33 \citep{mookerjea}.

The global $\Sigma_{SFR}(\Ha)$ calculated for these tidal tails is
two to three orders of magnitude lower than those found in spiral
galaxies as well as dwarf galaxies. \citet{kennicutt} find that $ -3.3
< \log(\Sigma_{SFR}) < -0.8$ in 61 normal spiral galaxies.  The Milky
Way galaxy has $\Sigma_{SFR} = 3.6 \times 10^{-3}$ \msun yr$^{-1}$
kpc$^{-2}$ \citep{naab}.  The LMC has a global star formation rate of
0.4 \msun yr$^{-1}$ which gives $\Sigma_{SFR} = 1.5 \times 10^{-3}$
\msun yr$^{-1}$ kpc$^{-2}$.  The SMC has a global star formation rate
of 0.05 \msun yr$^{-1}$ which gives $\Sigma_{SFR} = 1.9 \times
10^{-4}$ \msun yr$^{-1}$ kpc$^{-2}$ \citep{wilke}.  

The $\Sigma_{SFR} (gas)$ is similar in magnitude to values in the LMC
and SMC.  The Magellanic Stream
is a very nearby example of a gas tail of presumably tidal origin that has no 
star formation associated with it.  \citet{putman} measured the total {\HI} gas mass
in the Stream to be $2.1 \times 10^8$ \msun.  Converting the angular size of the Stream
($100 ^{\circ} \times 10 ^{\circ}$) using a distance of 55 kpc \citep{putman}, we 
infer a size of the Magellanic Stream to be 940.9 kpc$^2$.  Using the resulting gas density of 
$\Sigma_{\HI} =  2.2  \times 10^5$ \msun kpc$^{-2}$,
the Magellanic Stream has an expected $\Sigma_{SFR} = 3 \times
10^{-5}$ \msun yr$^{-1}$ kpc$^{-2}$ which is two orders of magnitude lower than that 
calculated for the Western tidal tail of NGC 2782 and four orders of magnitude lower than
the Eastern tail.  

The difference between the $\Sigma_{SFR}$ values calculated
from the {\Ha} flux and that predicted from the gas density may
indicate a lower star formation efficiency \citep{knierman12}.  However, the {\Ha} SFR
is a lower limit since it only traces massive star formation in the last 5 Myr.  Also, the {\Ha} emission
may depend on the masses of the star clusters formed.  If the tails
only formed lower mass star clusters where few high mass stars reside, similar to the Taurus-Auriga region \citep{kenyon}, there would be young, blue star clusters but a lack of widespread {\Ha} emission.  The
Western tail has only one small in size {\HII} region, but several other blue star cluster candidates. \emph{Galaxy Evolution Explorer} (\emph{GALEX}) All-sky Imaging Survey \citep[AIS;][]{galex} images show faint UV emission along the Western tail, indicating a young stellar population, probably dominated by B and A stars \citep{knierman12, tf12}.
In addition, the emission is concentrated only in specific areas in both tails rather
than being spread over the tail region.  This may indicate that global
star formation rates from {\Ha} or {\CII} are not an accurate assessment of the
total star formation in the tails.

\subsubsection{Star Formation on Local Scales}
We also examine the star formation on local scales in both tails using 
{\Ha} and {\CII}.  Table \ref{herschel} shows the SFR from {\Ha} for these massive {\HI} clumps using \citet{kennicutt} 
which range from 0.0016-0.008 \msun yr$^{-1}$ in the Eastern tail and 0.015 \msun yr$^{-1}$ for 
the Western tail source.  The Western tail source has a factor of 2-3 higher SFR than TDGC and E-HI-S the two most luminous {\Ha} sources in the Eastern tail.  
Using the {\CII} luminosity and the equation for SFR from \citet{boselli}, the range of SFR
in the Eastern tail is 0.017-0.035  \msun yr$^{-1}$ and $<0.02$  \msun yr$^{-1}$ in the Western tail.
The SFR is a factor of 7 larger from {\CII} than from {\Ha} for E-HI-S the most luminous {\CII} source in the Eastern tail while for TDGC,  SFR(\CII)$ =4$ SFR(\Ha).  We expect a lower SFR(\Ha) than SFR(\CII) since {\Ha} is affected by extinction and also only traces the most recent star formation.  That the SFR(\CII) is larger than {\Ha} may indicate that  star formation in the Eastern tail may proceed differently due to its origin as a "splash region".  
However, in the Western tail the {\HII} source has SFR(\Ha) similar to the upper limit for the SFR(\CII).  Since the Eastern tail shows a higher SFR(\CII) than SFR(\Ha), the at most equal values in the Western tail, but likely lower SFR(\CII) indicates that perhaps something  is happening in the Western tail to suppress {\CII} emission.  

These comparisons may be affected by differing areas, so we calculate the SFR per area ($\Sigma_{SFR}$) to compare regions.  The local values for SFR per area from  {\Ha}
 are $2-11 \times 10^{-4}$  \msun yr$^{-1}$  kpc$^{-2}$ for the regions in the Eastern tail.  The TDGC region has a similar $\Sigma_{SFR}$ to the Western tail {\Ha} region ($\Sigma_{SFR} = 1\times 10^{-3}$  \msun yr$^{-1}$  kpc$^{-2}$).  The remainder of the Eastern tail sources are 2-4 times lower than the Western tail source.  This seems to indicate local star formation as indicated by  {\Ha} is occuring at similar levels in each tail.  
The $\Sigma_{SFR}(\CII)$ in the Eastern tail ranges from $2.5-4.5 \times 10^{-3}$ \msun yr$^{-1}$  kpc$^{-2}$
and in the Western Tail is  $<1.1 \times 10^{-3}$ \msun yr$^{-1}$  kpc$^{-2}$.  The TDGC region has the highest  
$\Sigma_{SFR}(\CII)$ at  2 times the other two detected regions and this region is 4 times larger than the limit in the Western tail.

Using the total gas surface density, we find even higher expected star
formation rates.  In the Eastern tail,  we expect $\Sigma_{SFR}$ of 0.05-0.18 \msun
yr$^{-1}$ kpc$^{-2}$.  Due to the non-detection of CO in the Western
tail, the star formation rates from the gas density are upper limits.
Even so, these upper limits are lower than the Eastern tail by a 
factor of 10-40 ($\Sigma_{SFR}(gas) < 0.005-0.009$ \msun
yr$^{-1}$ kpc$^{-2}$). 

In summary, on local scales, we find the Western tail {\HII} region has the highest SFR using {\Ha} as a tracer, but has a low SFR using its upper limit in {\CII}.  This indicates that {\CII} is suppressed relative to {\Ha} in the Western tail.  
The Eastern tail has a higher SFR from {\CII} as compared to {\Ha} perhaps indicating that this "splash region" has a different star formation law than in normal galaxies.  
Normalized to area, we find similar SFR per unit area in both tails using {\Ha} as a tracer.  This seems to indicate that star formation is occurring locally at similar rates in both tails.  The expected SFR per area from the local gas density is higher than that observed by both {\Ha} and {\CII} indicating that there may be a lower SFE at the local level in both tails as well.  Comparing the tails, the expected local SFR surface density from the observed local gas density is 14-40 times higher in the Eastern tail than in the Western tail.  This indicates that there is a larger gas reservoir for star formation in the Eastern tail at least on local scales.

\subsection{Star Formation Efficiency}
Having calculated the SFR for each region, we now examine the star formation efficiency (SFE) for these local regions.    \citet{boquien} use multiwavelength data of major merger Arp 158 to study the local Schmidt-Kennicutt law.  They find that star forming regions in the tidal debris follow a different Schmidt-Kennicutt law than those in the central regions of the merger, falling along a line of similar slope to \citet{daddi}, but offset so that the same gas density gives lower values of SFR.  As discussed in \citet{knierman12}, W235 is consistent with the other star forming regions in the tidal debris of Arp 158 indicating a lower SFE than in the central region of Arp 158.  If we compare the Eastern tail star forming regions to the tidal debris of Arp 158, we find that for their given total gas surface density, their star formation rates from {\Ha} lie more than 2 dex below the relation for the tidal debris of Arp 158 indicating a substantially lower SFE in the Eastern tail.  

We calculate the gas depletion timescales ($\tau_{dep}$) or the amount to time it would take for the current SFR to deplete the gas (either molecular or atomic).  The SFE is the inverse of the gas depletion time.  
Table~\ref{tab:sfe} lists region, depletion timescale of molecular gas mass vs. SFR from {\Ha}  ($\tau_{\rm dep, H_2} = $M$_{\rm mol}/$SFR$_{\Ha}$), depletion timescale of neutral gas mass vs. SFR from {\Ha} ($\tau_{\rm dep, \HI} = $M$_{\rm HI}/$SFR$_{\Ha}$), depletion timescale of molecular gas mass vs. SFR from {\CII} ($\tau_{\rm dep, H_2} = $M$_{\rm mol}/$SFR$_{\CII}$), and depletion timescale of neutral gas mass vs. SFR from {\CII} ($\tau_{\rm dep, \HI}  =  $M$_{\rm HI}/$SFR$_{\CII}$)) for the local and global regions in both tails.  

The highest SFE ($\tau_{\rm dep, H_2} < 1.5$ Gyr) is in the Western tail {\HII} region considering the molecular gas limit and {\Ha} SFR.  This is comparable to the molecular gas depletion timescales determined for the star forming regions in Arp 158 \citep[$\tau_{\rm dep} \sim 0.5-2$ Gyr;][]{boquien} and in TDGs \citep[$\tau_{\rm dep} \sim 0.8-4$ Gyr;][]{braine}.  These ranges are also similar to the average gas depletion timescales in spiral galaxies.  If we use the value for the CO emission assumed in the discussion of \citet{tf12} using an analog {\HII} region in the tidal debris near NGC 3077, we might expect it to have molecular mass of $1.5\times 10^6$ \msun.  If we use this value for the molecular mass, $\tau_{\rm dep, H_2} = 0.1$ Gyr,  making this region similar to gas depletion timescales in dwarf galaxies (1 -100 Myr). If this molecular mass is close to the actual value, this would mean that this region is highly efficient at making stars.  

Considering neutral gas and SFR from {\Ha}, TDGC and W235 ($\tau_{\rm dep, \HI} = 5-7.7$ Gyr) are less efficient than normal star forming regions.  But these areas are still more efficient at star formation than outer regions of spiral galaxies at $r_{25}$ ($\tau_{\rm dep, \HI} \sim 20$ Gyr) or dwarf galaxies at $r_{25}$ ($\tau_{\rm dep, \HI} \sim 40$ Gyr).  Using the molecular gas and  {\Ha} SFR, TDGC and E-HI-S have depletion timescales of 33 Gyr and 38 Gyr similar to these less efficient regions.  \citet{bigiel} find very low SFE ($\tau_{\rm dep, \HI} \sim 100$ Gyr) in the outer disks of spiral galaxies using FUV and {\HI} observations.  
Even less efficient than outer disk regions is the E-HI-mid region. \citet{smith99} also found low star formation efficiency in the Eastern tail using the ratio of $L_{\Ha}/M_{H_2}$.  Overall, our results comparing SFR(\Ha) and gas mass indicate that on local scales the Western tail is more efficient at forming stars than the Eastern tail.  

Considering molecular gas and SFR from {\CII}, the Western tail limit remains with the highest SFE, however, both the gas mass and SFR(\CII) are upper limits.  In the Eastern tail, E-HI-S has the highest SFE, but with a depletion timescale of 5.6 Gyr is still less efficient than normal star forming galaxies.  Considering the neutral gas, TDGC has the highest SFE ($\tau_{\rm dep, \HI} = 1.3$) indicating efficient star formation.  Using {\CII}, the Western tail is also more efficient at star formation than most of the Eastern tail except for TDGC considering its {\HI} mass.

Since the CO observations were taken with a large beam, we examine the Eastern tail on larger scales.  Even if we consider the total molecular and neutral hydrogen for the entire Eastern tail, the SFE remains low (16-43 Gyr).  For the Western tail as a whole, we sum the limits from the three CO pointings and use the total {\HI} mass from \citet{smith94}.  The Western tail has a low SFE considering the molecular gas mass limits and the single {\HII} region ($<12$ Gyr), but is very inefficient considering the {\HI} mass of the tail (133 Gyr).  So on global scales, it appears that the Western tail is less efficient at forming stars.  

On both global and local scales, it appears the Western tail is more efficient at forming stars than the Eastern tail.  The Western tail also lacks high mass star clusters and star cluster complexes which are hosted in the Eastern tail.  These two points are not inconsistent.  In the Eastern tail, star cluster complexes of higher mass will form many more high mass stars than isolated lower mass star clusters in the Western tail.  These high mass stars provide feedback and energy into the ISM in the form of stellar winds and supernova.  This feedback can suppress further star formation and have a lower SFE in the area.  In addition, \citet{smith99} suggest that the Eastern tail formed as a ``splash" region (versus a tidal region of merger debris) and may have inhibited star formation due to gas heating during the encounter.  A tidally formed region such as the Western tail, would have gravitational compression and possibly enhanced star formation.  Given the higher SFE in the Western tail, this may be evidence for gravitational compression in the tidal tail.

\begin{deluxetable}{ccccc}
\tabletypesize{\scriptsize}
\setlength{\oddsidemargin}{-0.5in}
\tablecaption{Comparison of Star Formation Efficiencies Between Eastern and Western Tails}
\tablewidth{7.15in}
\tablewidth{0pt}
\tablehead{\colhead{Location} & \colhead{$\tau_{\rm dep, H_2, \Ha}$} & \colhead{$\tau_{\rm dep,{\rm \HI, \Ha}}$ }   &\colhead{$\tau_{\rm dep, H_2, \CII}$} & \colhead{$\tau_{\rm dep,{\rm \HI, \CII}}$ }  \\
\colhead{} & \colhead{(Gyr)} & \colhead{(Gyr)} & \colhead{(Gyr)} & \colhead{(Gyr)} }
\startdata
East & &\\
TDGC & 33 & 5 & 8.2 & 1.3\\
\HI-N & $81$ & 160 & $>$70 & $>$141\\
\HI-M & 232 &  200 & $21$ & $18$\\
\HI-S & 38 & 81 & 5.6 & 12\\
\hline
%FFE & 22 & 26 & 5 & 6\\
%SFFE & 32 & 70 & 7 & 14\\
E Tail & 16\tablenotemark{a} & 43 & 2.7\tablenotemark{a} & 7 \\
\hline
\hline
West & &  \\

\HI-N & $<$29\tablenotemark{a} & $>$240 & &  \\
\HI-M & $<$1.5& $7.7$ & $<$1.4\tablenotemark{b} & $>$7.3\\
\HI-S & $<$53\tablenotemark{b} & $>$390 & & \\
\hline 
W Tail &  $<12$ & 133 & $<4$\tablenotemark{b} & $>45$  \\

\enddata
\tablenotetext{a}{Using observations from this work.}
\tablenotetext{b}{Both gas mass and SFR are upper limits.}  

\label{tab:sfe}

\end{deluxetable}

%\clearpage

%
\section{Conclusions}
Tidal tails provide laboratories for star formation under extreme conditions
very different from quiescent galaxy disks. With low gas pressures and
densities and smaller amounts of stable molecular gas they are perhaps
on the edge of the parameter space open to star formation.  

The two tails of NGC 2782 are an interesting place to consider in this 
discussion.  The Western tail is rich in {\HI} gas, but CO is not
observed in the massive {\HI} knots in the tail, leading \citet{braine}
to conclude that {\HI} ``...has presumably not had time to condense into
H$_2$ and for star formation to begin."  This study finds that at least 10
star clusters have formed in the Western tail based on the overdensity of bright, blue
objects in the tail.  The Western tail also hosts a bright {\HII} region which was
explored in \citet{knierman12} while 6 additional young regions are found in FUV and NUV by \citet{tf12} and in {\Ha} by \citet{werk}. Clearly the lack of {\it observable} CO
does not guarantee the absence of recent star formation. It may play a
role, however, in the properties of star clusters forming therein since the Western tail
lacks high mass star clusters and star cluster complexes. 
The Eastern tail has a
dense knot of {\HI} and CO-rich gas at its base. Based on the overdensity calculations, at least 14 young star 
clusters are observed in this region of the tail.  The Eastern tail also hosts several {\HII} regions \citep{smith99}, of which we discover {\CII} associated with five of these regions.  
In contrast, the Western tail {\Ha} source shows only an upper limit on {\CII} contrary to expectations based on its bright {\HII} region and its {\HI} column density.  

%This suggests that a larger,
%stable reservoir of molecular gas encourages the formation of more
%massive clusters.  

We examined in turn the reasons for the differences between the two tails:  ambient pressure, gas phase, SFR, amount of gas available for star formation, and efficiency of star formation.  

{\bf Ambient pressure:} Based on the calculation for $M_{char}$ in local regions in the Western tail, we find that the $M_{char}$ in the Eastern tail is only slightly larger than in the Western tail.  We surmise that the relatively small difference between the predicted clump mass in the two tails indicates that the ambient pressure between the two tails is not very different.  However, there is a large difference between $M_{char}$ and the upper limit on molecular mass in the Western tail indicating that the Western tail is deficient in molecular mass (or at least, in CO).  

 {\bf Gas phase:}  We find that the Western tail has much less molecular gas mass (traced by CO) for its neutral hydrogen mass than the Eastern tail.  This indicates that CO may be suppressed in the Western tail.  
Both tails are lower in their {\CII} intensity based on their {\HI} as compared to observations of nearby galaxies and Galactic regions.  The Eastern tail {\HI} regions are close, but still below, the expected {\CII} values for ``standard" {\HI} clouds while the Western tail limit is a few times below that.  We also find a lower $I_{\CII}/I_{CO}$ in the Eastern tail than studies of nearby star forming galaxies and Galactic {\HII} regions which is in line with GMCs and normal galaxies.  
% While the very different beam sizes of the CO and {\CII} observations may account for some of the discrepancy,  we are also exploring a different regime of star formation in these tidal tails for the first time.  

{\bf SFR:} Using the entire tail area, we find that the global SFR from {\Ha} and from {\CII} are much less than the SFR expected from the gas surface density suggesting that both tails have low SFE.  On local scales, we find the Western tail {\HII} region has the highest SFR from {\Ha}, but has only an upper limit for SFR from {\CII}.  This indicates that {\CII} emission is suppressed in the Western tail relative to {\Ha}.  Normalized to their areas, we find similar local $\Sigma_{SFR}({\Ha})$ in both tails indicating that star formation is occurring similarly on the local scale in both tails.  
For each local region, the expected SFR from the gas density is higher than from {\Ha} or {\CII} indicating lower SFE at the local level as well.  Comparing the two tails, the expected local SFR density from the gas density in the Eastern tail is 14-40 times higher than in the Western tail.  

{\bf Gas Reservoirs for SF:} We examined different tracers for the reservoirs of gas for star formation in these two tidal tails and found that these tracers were not consistent.  Both tails have abundant neutral hydrogen gas, but the Eastern tail has more molecular gas (traced by CO) and ionized gas (traced by {\CII}).  However, the Western tail has a higher local SFR and both tails have similar SFR per unit area, indicating that star formation according to {\Ha} emission is similar in both tails.  If we use {\CII} emission as a SFR tracer, we find that the Eastern tail has a higher SFR than the Western tail.  Looking at gas density, the Eastern tail has a higher total gas surface density than the Western tail.  So we expect a higher SFR per area in the Eastern tail. 
That these tracers are not consistent between the tails indicates that something different is going on between the tails.  In the Western tail, the non-detections in CO and {\CII} compared to its {\HI} and {\Ha} may be due to its inability to produce detectable CO or {\CII} emission.  This may be due to a deficiency in carbon even though it has a high oxygen abundance, because the tail material is going through its first generation of stars and has not had enough time to build up a higher carbon abundance.

{\bf SFE:}  By calculating the gas depletion timescales for molecular gas and neutral hydrogen, we examine SFE in the tails.  As discussed in \citet{knierman12}, the Western tail {\HII} region has a normal SFE when considering its molecular mass upper limit, but less efficient star formation if we consider its neutral hydrogen mass.   The Eastern tail regions have a low SFE, in some cases as low as spiral or dwarf galaxies at $r_{25}$.  We suggest that the lower SFE in the Eastern tail may be due to its more massive star clusters providing feedback to prevent further star formation.  Also, the Eastern tail may have a lower SFE due to its formation mechanism as a splash region where gas heating has a higher effect.  The Western tail has a higher SFE even though it has a lower local gas surface density than the Eastern tail, possibly due to its tidal formation where gravitational compression has more of an effect and can increase star formation.

%Consistent with previous observations, the Eastern tail of NGC 2782 has a
%population of young, star forming clusters with ages $< 200$ Myr.
%Also, we find that the Western tail has a population of young, star
%forming clusters with ages $< 200$ Myr indicating that star formation
%has occurred within this tail.  This is contrary to speculation that the
%lack of molecular gas in the Western tail means that no stars were
%forming there.  
%The Western tail lacks high mass young star clusters.  We examined whether this was due to its lower
%ambient pressure, gas phase, gas surface density, or star formation efficiency.  

%

\begin{acknowledgments}
We acknowledge helpful discussions with J. Lee, S. Malhotra, P. Young, J. Monkiewicz, and J. Funes.  We wish to thank R. Kennicutt for use of his {\Ha} filter.  We thank the anonymous referee for very helpful suggestions which have improved this paper.
KK acknowledges support by the University of Arizona/NASA Space
Grant Graduate Fellowship (2004-2006) and also by NASA through an award issued by JPL/Caltech.  ISK is the recipient of a John Stocker Postdoctoral Fellowship from the Science and Industry Research Fund.  
We used the NASA/IPAC Extragalactic Database (NED) which is operated by JPL, California Institute of Technology, under NASA contract.   We acknowledge the usage of the HyperLeda database (http://leda.univ-lyon1.fr).  This work is based on observations with the \emph{Vatican Advanced Technology Telescope} (\emph{VATT}): the Alice P. Lennon Telescope and the Thomas J. Bannan Astrophysics Facility, with Herschel, a European Space Agency Cornerstone Mission with significant participation by NASA, and with the Kitt Peak 12 Meter which is operated by the Arizona Radio Observatory (ARO), Steward Observatory, University of Arizona.  
\end{acknowledgments}

%\clearpage

\end{document}